\documentclass[psfig,aps,12pt,floatfix]{revtex4}

\begin{document}

\title{Dynamics of membranes driven by actin polymerization.}
\author{Nir Gov}
\address{Department of Chemical Physics,
The Weizmann Institute of Science,\\
P.O.B. 26, Rehovot, Israel 76100}
\author{Ajay Gopinathan}
\address{Department of Physics and Materials Research Laboratory,\\
University of California Santa Barbara,\\
Santa Barbara, CA 93106-9530 ,U.S.A.}

\begin{abstract}
A motile cell, when stimulated, shows a dramatic increase in the activity 
of it's membrane, manifested by the appearance of dynamic membrane structures 
such as lamellipodia, filopodia and membrane ruffles. The external stimulus turns on membrane bound activators, like Cdc42 and PIP2, which cause
increased branching and polymerization of the actin cytoskeleton in their vicinity 
leading to a local protrusive force on the membrane. The emergence of the complex
membrane structures is a result of the coupling between the dynamics of the membrane, 
the activators and the protrusive forces. We present a simple model that treats the dynamics of a membrane under the action of actin polymerization forces that depend on the local density of freely diffusing activators on the membrane. We show that, depending on the spontaneous membrane curvature associated with the activators, the resulting membrane motion can be wave-like, corresponding to membrane ruffling and actin-waves, or unstable, indicating the tendency of filopodia to form. Our model also quantitatively explains a variety of related experimental observations and makes several testable predictions.
\end{abstract}

\maketitle

\input epsf 

\section{Introduction}

Various types of directed cell motility are driven by the
polymerization of an actin network that exerts a force on
the cell membrane, pushing it forward. During cell motility, the
leading edge of the cell exhibits a range of dynamic structures
such as lamellipodia, filopodia and membrane ruffles
(Small,2002;Pollard,2003). These dynamic surface patterns of
moving cells are usually observed to have lengthscales in the
$1\mu$m range, and appear in many different cell types
(Gungabissoon,2003). The lamellipodium is a flat, disk-like
extensional structure generally occurring at the periphery of
spreading cells while the filopodia or micro-spikes are actin rich
needle-like structures seen generally as extensions of the
lamellipodium. To generate movement, the cells use precursor
contacts found in membrane ruffles, or on the underside of
filopodia, which can help form adhesive contacts. The lamellipodia
are generally extruded in the direction of a strong signal such as
a chemoattractant that induces cell migration. The extracellular stimulus turns on certain membrane bound activators that in turn activate a series of  proteins that trigger actin branching and polymerization leading to a directed and regulated protrusive force. The interplay between the dynamics of the activators, the protrusive forces generated by the actin polymerization and the membrane dynamics results in the rich variety of dynamic structures described above.  This kind of actin based motion
is ubiquitous with examples ranging from the chemotaxis of
macrophages to the movement of metastatic tumor cells. Since cell
motility depends so crucially on the formation of these dynamic
membrane structures, it is imperative to understand the origin and
dynamics of these structures.

There have been a number of theoretical approaches to the problem
of cell motility driven by actin polymerization
(Mogilner,1996;Bottino,2002;Grimm,2003;Stephanou,2004). These
studies focus on the inter-relationship between the dynamics of
actin polymerization and the protrusive forces generated which
lead to propulsion. However, one important aspect that has been
neglected so far is the crucial role played by membrane bound activators (henceforth referred to simply as activators or membrane proteins), and  in particular the thermal density
fluctuations and the spontaneous membrane curvature associated
with the activators. Activators that have been well
studied include the Rho family GTPase Cdc42 and the membrane
phospholipid PIP2 (Higgs,2001). These activators bind to, and
activate, WASp/Scar family proteins by inducing a conformational
change. The WASp/Scar proteins in turn activate Arp2/3 which is
directly responsible for generating new branches in the actin
network (Higgs,2001;Blanchoin,2000;Carlier,2003). Fluctuations in
the density of the membrane activators can thus directly lead to
fluctuations in the actin branch density and hence the protrusive
force. Also important is the fact that both Cdc42 and PIP2 have
been shown to induce and sense membrane curvature by binding to
BAR domain proteins and epsin respectively
(Habermann,2004;Ford,2002). In this paper we propose an
approach that takes into account density fluctuations and membrane
curvature associated with the membrane activators. The dynamics of
the actin polymerization, and its dependence on the relative
concentration of actin and supporting proteins, has been
calculated by Carlsson (Carlsson,2003). This gives us
the steady-state velocity of the advancing membrane and actin gel,
as a function of the branching rate which is directly proportional
to the membrane density of the activators. We therefore separate
the in-cell dynamics of the actin polymerization, from the
in-membrane dynamics of the activators. The diffusion and
spontaneous curvature associated with these membrane activators
will determine the time and length-scales of the dynamic patterns.
This allows us to write a simple model and arrive at analytical
expressions, while still preserving the rich variety of dynamical
behavior that is observed. The merits of such an approach, apart from
the knowledge gained concerning dynamical models, lie in its
quantitative and testable predictions for cell motility in vivo.

Our work draws on previous models of instabilities in active
membranes
(Chen,2004;;Sankararaman,2004;Manneville,2001;Ramaswamy,2000;Prost,1998),
some of which we have combined here into a simpler
form. The previous analyses are different in essential ways
from the present study. They consider the case where the active membrane
proteins are ion-pumps and therefore solve for the fluid
circulations, which is not necessary here. The first analysis
(Chen,2004) considers a system close to a phase separation
transition, which is not a constraint in our work. The possibility
of wave-like propagation for negative spontaneous curvature of the
membrane proteins, was also not considered
(Chen,2004;Sankararaman,2004). The second analysis
(Manneville,2001;Ramaswamy,2000) does find propagating modes, but
does not specifically relate them to the spontaneous curvature of
the membrane activators. Our model is therefore an application of
these previous studies to the case of actin-driven cellular
motility. It allows us to describe, in a very detailed and
transparent way, the physics of this system in terms of an active
membrane.

\section{Model}
We now introduce our model for the dynamics of the coupled system consisting of the membrane, the activators and the actin polymerization induced forces. 
The problem we wish to solve is shown schematically in Fig.1. We
have an average areal density of activating proteins, $n_0$, which
induces actin polymerization such that the membrane moves forward
at a velocity given by: $v_0=An_{0}\sim1-0.1\mu m$/sec, where $A$
is a coefficient that depends on various factors, such as
concentration of actin monomers, temperature etc., and
$n_{0}\sim2\cdot10^{15}$m$^{-2}$. The time evolution of the density of these activating proteins is described by
a diffusion equation (Manneville,2001;Ramaswamy,2000)
\begin{equation}
\frac{\partial n}{\partial t}= D \nabla^2n-\Lambda \kappa
\overline{H}\nabla^4 h+\nabla\cdot f_n \label{diffusion}
\end{equation}
where $D\sim 1\mu m^2/$sec is the in-membrane diffusion
coefficient of the proteins, $\kappa$ is the bending
modulus of the membrane (typically $\kappa\sim10k_{B}T$),
$\Lambda=D/\chi$ is mobility of the proteins in the membrane where 
$\chi\simeq k_{B}T$ is the effective in-plane compression energy
of the proteins, $\overline{H}$ is the spontaneous curvature of
the membrane proteins, and the thermal noise force satisfies the
following correlation: $\langle f_{n}(r,t)f_{n}(r',t')
\rangle=2n_{0} \Lambda k_{B}T\delta(r-r')\delta(t-t')$. The first term describes the free diffusion of the activators on the membrane, while the last term captures the effect of thermal noise. Evidence
for free diffusion of the membrane proteins that activate the
polymerization of actin, appears in (Gerisch,2004). The second term takes into account the coupling between the spontaneous curvature of the activators and the local curvature of the membrane. Here $h$ refers to the coordinate that measures the normal displacement of the membrane from a flat reference plane.

The membrane deviation from flatness obeys the following equation of motion
\begin{equation}
\frac{\partial h}{\partial t}=-\int dr'\Lambda(r-r')\kappa\nabla^4
h(r')+An \label{eqmotion}
\end{equation}
for a free, flat membrane. Here the first term is simply the response of  a membrane that is surrounded by a fluid, characterized by the hydrodynamic interaction kernel $\Lambda$. After Fourier transforming into
$q$-space, the hydrodynamic interaction kernel is given by 
$\Lambda(q)=1/4\eta q$, and the response of the free membrane is
 $\omega_{q}=\kappa q^3/4 \eta$, where $\eta$ is the
viscosity of the surrounding fluid. We note that if the edge of
the membrane is highly curved, then the membrane response is
different. We take the response in this limit to be: $\omega_{q,1}\sim \kappa
q/4\eta d^2$, where $d$ is the local radius of curvature at the
membrane edge (Fig.1), assuming that the hydrodynamic interaction
(Oseen kernel) remains the same as that for the flat membrane (Zilman,2002). A
similar form for the response of a flat membrane occurs when 
tension is dominant and is given by  $\omega_{q,t}=\sigma q/2 \eta$, where
$\sigma$ is the effective surface tension. This regime occurs for
wavevectors: $q<q_{t}=\sqrt{\sigma/\kappa}$. In the analysis below,
the results using either $\omega_{q,1}$ or $\omega_{q,t}$ are
interchangeable, by making the transformation: $\kappa/2d^2\leftrightarrow \sigma$. The second term describes the action of the actin polymerization induced forces, whose effect we model by the addition of a velocity to the membrane. This velocity is taken to be directly proportional to the local density of activators, with a proportionality constant, $A$, as described above. The two equations (eqs. \ref{diffusion},\ref{eqmotion}) form a coupled set that completely describes the dynamics of our system. We now describe, in detail, the assumptions that we make in our model.

\subsection{Discussion of assumptions}
In this paper, we assume that the membrane proteins do not
bind to the actin network, and that the diffusion coefficient (in eq.\ref{diffusion}) is
homogeneous. The diffusion coefficient, $D$, may be treated as an
effective value, which takes into account the average (uniform)
effect of actin interactions with the membrane proteins. This is
an approximation, since the diffusion coefficient is likely to 
decrease when the density, $n$, increases, due to a ``crowding
effect'' (Almeida,1995). It may also depend on the actin density.
Future work may include the dependence of the
diffusion coefficient on the actin density, membrane curvature and
the dynamics of protein-actin binding/unbinding. The binding of
the membrane proteins to the actin network that they nucleate,
introduces an effective attraction between them, which may drive
phase separation. An effective protein-protein interaction will
introduce a term of the form: $J(r-r') n(r)n(r')$, to
Eq.\ref{diffusion}. This possibility is deferred to future
studies.

We use the term ``activator-proteins'' in the most general
sense: it stands for membrane proteins that trigger actin
polymerization, branching or bundling, that produces in turn a
protrusive force on the membrane. In addition to the
activators mentioned above, there are, for example, the VASP
membrane proteins that recruit fascin, which
cross-links actin filaments into bundles. These bundles can push more
effectively on the membrane, and produce filopodia (Biyasheva,2004).
We consider a single species of membrane activator, which is also
constantly in its ``on'' state. A more detailed description could
allow for the kinetics of the turning-on of these activators,
which is a process influenced by the presence of other proteins,
such as external chemotactic signals or binding of other cellular
proteins, the average density $n$ and membrane curvature.
Nevertheless, in this paper, we wish to investigate the
dynamics that arise from the simplest model first.

In Eq.\ref{eqmotion}, we neglect the thermal fluctuations of the
membrane, which are usually much smaller than the motion due to
the actin polymerization (see Section IV). The thermal
fluctuations of the membrane and the
actin-induced motion described above are incoherent (decoupled) , so that they simply add
to the overall mean-square height fluctuations. We further
expect the thermal fluctuations of the membrane to be almost
eliminated when the membrane is being pushed by the actin network,
since any membrane motion that is not
synchronized with the actin polymerization, such as the thermal
motion, will be reduced to negligible values due to the large
bulk modulus $Y$ of the actin gel, giving mean-square height fluctuations: $\langle
h^2\rangle\propto k_{B}T/Y$. The membrane response, $\omega_{q}$,
which we used above, for a membrane that is pushed by the actin
polymerization, describes the dynamics of the fluid flow outside
the cell, and does not need to include the elasticity of the actin
network. For the same reason, membrane motion that arises
directly from the spontaneous curvature of the membrane proteins
(Chen,2004;Manneville,2001;Ramaswamy,2000;Sankararaman,2004), is
also negligible since it too is incoherent with respect to the motion due to actin polymerization.

In Eq.\ref{eqmotion} we also assume that there is a linear relation
between the density of  activators, $n$, (and therefore of activated
Arp2/3 protein) and the forward velocity of the membrane. That this
assumption is valid for low densities and velocities (i.e.
velocities low with respect to the saturation polymerization
velocity: $v_{p}\sim 1\mu$m/sec) has been shown explicitly within
the context of a model that considers an obstacle driven forward
by a polymerizing  actin network with a spatially homogeneous
branching rate (Carlsson,2003). Above a very small cut-off
branching rate, the forward velocity increases linearly with the
branching rate and saturates to a maximum at high branching
rates. This behavior is generic and not crucially model dependent.
One can also reach the same conclusion from a continuum
perspective. At low densities the branching of the actin gel increases
linearly with the activator density (Blanchoin,2000). This means that the bulk modulus of the gel will aslos be a linear function of the activator density: $Y\propto n$. The relation between the velocity
normal to the plane of the membrane and the modulus of
the pushing gel is given by (Gerbal,2000)
\begin{equation}
\frac{v}{v_{p}}=\frac{1}{1+\frac{F_{ext}}{Y S_{b}}} \label{gel}
\end{equation}
At low gel densities, Eq.\ref{gel} implies: $v/v_{p}\simeq Y
(S_{b}/F_{ext})$, where $F_{ext}$ is the external drag force and
$S_{b}$ is the local area of the membrane which is pushed by the
actin gel. Thus we do find a linear relation between the velocity
and the density $n$. At high densities the velocity saturates at
$v_{p}$, and our analysis breaks down.

Note that our description of an imposed velocity, $v(n)=An$ in
Eq.\ref{eqmotion}, due to the actin polymerization, is different
from that of an imposed force condition $F_{actin}$
(Manneville,2001;Ramaswamy,2000;Gov,2004). The imposed velocity
condition is natural if it is determined by the dynamics of the
actin polymerization in the lamellipodia. This condition also
applies if there is a roughly constant drag force due to the actin
gel itself (Gerbal,2000). On the other hand, if the motion is determined by the
action of the drag force of the surrounding fluid, then it is more
natural to keep the imposed force condition, and substitute:
$A\rightarrow F_{actin}n/\eta q$, in Eq.\ref{eqmotion}.

Another assumption implicit in Eq.\ref{eqmotion} is that changes
in the membrane density of the activators translates
instantaneously into changes in the force with which the actin gel
is pushing the membrane. This is not strictly true, and we now
wish estimate the time lag for this process. First there is the
chemical time for Arp2/3 activation. This is of order $1$msec,
which translates to membrane density fluctuations of length-scale
$10$nm, which, in turn, is much shorter than the typical mesh size of
the actin gel ($\sim50$nm). So we may neglect this contribution to the time lag.  There is growing evidence that the new
branches are formed directly at the free barbed ends of the actin
filaments, that are in contact with the membrane (Carlier,2003).
In this case, there would be no other source of time lag. However,
if the Arp2/3 can diffuse into the bulk and nucleate new branches at
barbed ends further back from the leading edge, it would be
another source of time-lag. Experimental observations suggest that
barbed ends are localized to a width of $\lambda\sim 100$nm from
the leading edge (Bailly,1999). The average time for a new
filament to grow and cover that distance back to the plane of the
membrane, and add to the pushing force, is:
$t_{\lambda}=\lambda/v_{p}\sim300$msec. On this time scale, the
membrane density fluctuations can diffuse away over length-scales
smaller than $\sqrt{t_{\lambda} D}\sim200$nm. This length is of the order of
4 unit mesh sizes of the actin gel, while we are interested in a
continuum description that is valid over longer length scales. We
therefore conclude that, within these limitations, for length-scales
longer than 200nm, we can neglect the time-lag.

\section{Results}
We now solve for the dynamics of the system by first fourier transforming both Eqs. (\ref{diffusion},\ref{eqmotion}),
and using solutions of the form: $e^{-i(\omega t+q\cdot r)}$. This gives the following system of equations in matrix form
\begin{equation}
\left(%
\begin{array}{cc}
  -i \omega+\omega_{D} & -B q^4 \\
  -A & -i \omega+\omega_{q} \\
\end{array}%
\right)\left(%
\begin{array}{c}
  n \\
  h \\
\end{array}%
\right)= \left(%
\begin{array}{c}
  -i f_{n} q \\
  0 \\
\end{array}%
\right)\label{matrix}
\end{equation}
where $B\equiv \Lambda \kappa \overline{H}$ and $\omega_{D}=D
q^2$. The dispersion equation of the protein density $n$ and
membrane height $h$ respectively, is given by equating the
determinant of the matrix in Eq.\ref{matrix} to zero, which yields
\begin{eqnarray}
\omega_{n}&=&-i\frac{1}{2}\left(\omega_{D}+\omega_{q}-\sqrt{4ABq^4+\omega_{D}^2-2\omega_{D}\omega_{q}+\omega_{q}^2}\right)
\nonumber \\
\omega_{h}&=&-i\frac{1}{2}\left(\omega_{D}+\omega_{q}+\sqrt{4ABq^4+\omega_{D}^2-2\omega_{D}\omega_{q}+\omega_{q}^2}\right)\label{omega}
\end{eqnarray}
where, the time-dependent response is given by $(n,h)(t)\propto
\exp{(-i\omega_{n,h} t)}$. It is to be noted that these solutions decay exponentially in time if $Re[\omega_{n,h}]>0$. We now discuss the results for two different cases, positive ($\overline{H}>0$) and negative ($\overline{H}<0$)
spontaneous curvature of the membrane activator proteins.

\subsection{Positive spontaneous curvature}

Membrane activators with positive spontaneous curvature
($\overline{H}>0$) will prefer to aggregate at the locations
with maximum local curvature (Fig.2). The solution for the membrane
height $h$, in this case, decays with time ($Im[\omega_{h}]>0,Re[\omega_{h}]=0$), while for the membrane
density of the activator proteins, $n$,  we find that there can be an
instability in the form of an exponentially increasing function of time ($Im[\omega_{n}]<0,Re[\omega_{n}]=0$)(Fig.3a). Depending on the
form of the membrane response we choose, we get unstable behavior for the
following range of $q$-wavevectors
\begin{eqnarray}
\omega_{q}&:& q<q_{c}, \quad q_{c}=\frac{4\eta AB}{\kappa
D}=\frac{4\eta v_{0} |\overline{H}|}{n_{0} \chi}
\nonumber \\
\omega_{q,1}&:& q>q_{c,1}, \quad q_{c,1}=\frac{\kappa D}{4\eta d^2
AB}\quad \textrm{or} \quad \frac{\sigma D}{2\eta AB} \label{qc}
\end{eqnarray}
These results arise when the bare response of the
protein diffusion is faster than the response of the membrane. In
this case, the protein aggregates in response to the membrane
curvature fluctuations, and builds up density fluctuations as
it responds faster (Fig.2). Similar instabilities due to
aggregation of membrane proteins with positive
spontaneous curvature were described in previous studies of
different active membranes
(Chen,2004;Manneville,2001;Ramaswamy,2000). In Fig.3a we plot
$\omega_{n}$ and $\omega_{h}$ as a function of $q$ (for the case
of a free tension-less membrane).

The general form of the instability criterion, $q<q_{c}$
(Eq.\ref{qc}), follows from comparing the time-scales of membrane
motion and in-membrane diffusion of the combined shape-density
undulations (Fig.2a). These undulations combine a local increase
in the membrane protein density, with the driven (active) normal
motion of the membrane. The motion of these shape-density
undulations can be described by an effective diffusion with a dispersion relation given by 
$q^2=\omega_{bump}/D'$, where $D'=AB/D=v_{0} |\overline{H}|
\kappa/n_{0} \chi$. We now discuss the parameters that control this motion.   The membrane ``bump'' diffuses faster when the driving
velocity produced per membrane protein, proportional to
$v_{0}/n_{0}$, is larger. This is because, density fluctuations are
converted faster into a height undulation. Larger spontaneous
curvature, $\overline{H}$, causes the membrane bumps to have smaller
wavelengths, which move faster. Finally, larger osmotic
pressure of the membrane proteins $\chi$ results in  a larger
wavelength of the density fluctuations, resulting in slower
motion. The unstable regime occurs for wavevectors where the
membrane response, $\omega_{q}$, is slower than the rate of
diffusion of density-bumps, $\omega_{bump}$. In this regime, the
aggregation of the membrane proteins can occur before membrane
undulations decay away. The criterion
appearing in (\ref{qc}) is simply a restatement of this result. It is to be noted that the final expression for $q_{c}$ (Eq.\ref{qc}) is independent of both the membrane bending modulus, $\kappa$, and the membrane protein
diffusion coefficient, $D$. Quantitatively, using the parameters
given before (see section II), we find: $D'\sim D/2$.

The instability we describe above does not lead to a real divergence,
since the velocity saturates at $v_{p}$ and the local
density of the membrane proteins saturates due to their finite
size. Note that a similar behavior of the critical wavevector of the membrane instability, was shown in
(Stephanou,2004): $k_{c}\propto k_{a}/D_{a}$, where $k_{a}$ is the
rate of actin polymerization, and $D_a$ is the bulk diffusion
coefficient of actin monomers in the cell cytoplasm. Comparing to
our expression for $q_{c}$ (Eq.\ref{qc}), we see that both results are
directly proportional to the rate of actin growth ($k_{a}$ or $v_{0}$) and
inversely proportional to the diffusion coefficient in the plane
of the membrane, which tends to smooth away the density
accumulation (in (Stephanou,2004) the actin diffusion was assumed
to be confined to a sub-membrane layer). The details of the two
models are nevertheless very different.

\subsection{Negative spontaneous curvature}

For negative spontaneous curvature of the membrane proteins
($\overline{H}<0$), we find that there is a range of wavevectors
for which the response frequencies $\omega_{n}$ and $\omega_{h}$
are real (Fig.3b). This corresponds to a wave-like behavior,
though still damped (or even over-damped) (Fig.2b). The range of
the wavevectors over which this occurs is given by the condition that the square-root term
in Eq.\ref{omega} becomes negative. This gives
\begin{eqnarray}
\omega_{q}&:& q<q_{w}, \quad
q_{w}=\frac{4\eta(2\sqrt{D'D}+D)}{\kappa}
\nonumber \\
\omega_{q,1}&:& q>q_{w,1}, \quad q_{w,1}=\frac{\kappa}{4\eta d^2
(2\sqrt{D'D}+D)} \label{qw}
\end{eqnarray}

The real parts of the frequencies for the height function and the activator density function are in anti-phase with each other
(Fig.3b), and correspond to an effective propagation velocity
$v_{eff}=Re[\omega]/q$ in the limit of small wavevectors
($q\rightarrow 0$)
\begin{eqnarray}
\omega_{q}&:& \quad v_{eff}=\frac{1}{2}q\sqrt{(4D'+D)D}
\label{vwave}
\end{eqnarray}
The damping of these waves is given in this limit by the factor:
$e^{-Dq^2t/2}$, coming from the membrane protein diffusion. For the
tension-dominated (second case in Eq.\ref{qw}), there are
propagating waves in the limit of large wavevectors ($q\rightarrow\infty$),
where we get
\begin{eqnarray} \omega_{q,1}&:& \quad
v_{eff,1}=\frac{1}{2}q\sqrt{(4D'+D)D} \label{vwave1}
\end{eqnarray}
and the damping of these waves is given in this limit by the
factor: $e^{-(Dq^2+\sigma q/2 \eta)t/2}$. The waves therefore
decay over a length-scale given by: $l\simeq
q^{-1}\sqrt{(4D'+D)D}/D$, which is of the order of the wavelength
of the density-height perturbation.

In Fig.3b we plot the imaginary and real parts of $\omega_{n}$ and
$\omega_{h}$ as a function of $q$ (using $\omega_{q}$ of the flat
and tension-less membrane). Note that for $q<q_{w}$ the imaginary
parts of both frequencies are equal, while the real parts have the same magnitude but opposite signs. For any choice of parameters, the motion of $n$
and $h$ changes from damped ($Re[\omega]>Im[\omega]$) to
over-damped ($Re[\omega]<Im[\omega]$) wave-like propagation as $q$
increases (see inset of Fig.3b). Finally, when $q>q_{w}$, we find
the usual exponential decay for both functions
($Re[\omega]=0,Im[\omega]>0$). At the critical wavevector $q_{w}$,
both the response frequencies have the value:
$\omega_{h}=\omega_{n}=-i16(\sqrt{AB}+D)(2\sqrt{AB}+D)^2\eta^2/\kappa^2$.

In the regime of wave-like propagation $q<q_{w}$ (or
$q>q_{w,1}$), the density fluctuations travel faster than the bare
diffusion, due to the additional curvature driving force
(Eqs.\ref{vwave},\ref{vwave1}) (Fig.3b). The driving force for the
wave-like propagation of density-curvature fluctuations is shown schematically in Fig.2b. A local increase in the protein density will result in
increased membrane curvature there, which then drives these
proteins into lower density areas due to their negative
$\overline{H}$, in addition to the usual diffusion. This
curvature-induced restoring force gives rise to the (albeit
damped) oscillatory behavior.

\subsection{Protein density and membrane height correlations}

So far, we have described the wavevector regimes where, depending on the sign of the spontaneous curvature of the activators, one gets either an instability or wave-like modes. We now wish to address the question of the actual amplitudes of the fluctuations that characterize the motion in these regimes. This we do by
 calculating the correlation functions of the membrane height and of the membrane activator density. Solving
Eq.\ref{matrix} we get
\begin{eqnarray}
\langle n^2(q,\omega)\rangle &=&\frac{\langle
f_{n}^2(q,\omega)\rangle q^2
(\omega^2+\omega_{q}^2)}{\omega^2(\omega_{D}+\omega_{q})^2+(\omega_{D}\omega_{q}-\omega^2-A
B q^4)^2}
\nonumber \\
\langle h^2(q,\omega)\rangle &=&\frac{\langle
f_{n}^2(q,\omega)\rangle q^2
A^2}{\omega^2(\omega_{D}+\omega_{q})^2+(\omega_{D}\omega_{q}-\omega^2-A
B q^4)^2}
 \label{corrfunc}
\end{eqnarray}
Integrating these functions over $\omega$ we find the spatial
(static) correlations $\langle n^2(q)\rangle,\langle
h^2(q)\rangle$. We plot these functions in Fig.4, for a free and
tension-less membrane. Analytic expressions for these functions
can be calculated, but are quite lengthy, so we will give them
explicitly only for the limiting cases.

It is to be noted that the height fluctuations in our model are derived solely
from thermally-driven density fluctuations of the membrane
proteins. This is why we get: $\langle
h^2(q,\omega)\rangle\propto\langle f_{n}^2(q,\omega)\rangle\propto
k_{B}T$. These height undulations are superimposed over the
average forward motion of the membrane, at average velocity
$v_{0}$, determined by the average density $n_{0}$.

Another notable point is that we assumed a continuous and constant force, or driving velocity, due
to the actin polymerization in Eq.\ref{eqmotion}. This is
reasonable as long as we are interested in membrane motions on
timescales longer than the duration of an individual actin
polymerization event. More generally, we can describe the actin-induced
velocity (or force) by a random shot-noise behavior (Gov,2004),
with a typical time $\tau$. This amounts to replacing:
$A^2\rightarrow A^2/(1+(\omega \tau)^2)$ in the numerator of
Eq.\ref{corrfunc} for $\langle h^2(q,\omega)\rangle$. This is
easily calculable, and is found to change the behavior
quantitatively, but not to change the value of the critical
wavevector $q_{c}$, or the qualitative forms of $\langle
h^2(q)\rangle$ in the $q\rightarrow0,\infty$ limits. 

We now discuss the form of the density and height correlations in various cases and limits.

For positive spontaneous curvature of the membrane proteins
($\overline{H}>0$) (Fig.4a), we find a divergence of both the
density and membrane height fluctuations at the critical
wavevector $q_{c}$ (Eq. \ref{qc}). Around the critical wavevector,
setting $q=q_{c}+\delta$, the divergences have the form
\begin{eqnarray}
\langle n^2(q_{c})\rangle&=&\frac{\langle
f_{n}^2\rangle}{2D}\frac{q_{c}^2
\kappa}{4\eta|\delta|(D+D')}=\frac{4\eta v_{0}^2 \overline{H}^2
k_{B}T\kappa}{\chi^3  n_{0}(D+D')}\frac{1}{|\delta|}
\nonumber \\
\langle h^2(q_{c})\rangle&=&\frac{4 \eta\langle f_{n}^2\rangle
A^2}{\sqrt{2}q_{c}^4 \kappa|\delta| D(D+D')}=\frac{k_{B}T}{\kappa
q_{c}^4}\frac{4 \eta v_{0}^2}{\sqrt{2}\chi n_0 (D+D')}
\frac{1}{|\delta|}
 \label{divcorrfunc}
\end{eqnarray}
where, from the last line, we can define: $\langle
h^2(q_{c})\rangle=k_{B}T_{eff}/\kappa q_{c}^4$. The
``effective-temperature'': $T_{eff}/T=4 \eta v_{0}^2/\sqrt{2}\chi
n_0 (D+D')\delta$, diverges at the critical wavevector. This is
 reminiscent of the divergence in the effective
temperature describing the response of active hair bundles in the
hair-cells of the auditory system (Martin,2001) when there is a resonance with an internal driving frequency. In our case the
divergence occurs when the length-scale of the active process is
in ``resonance'' with the length-scale of the spontaneous curvature.

In the limit $q\rightarrow0$, the height correlations have the form: $\langle
h^2(q)\rangle=k_{B}T_{eff}/\kappa q^4$. Here we chose to define an effective temperature, $T_{eff}$, since the power law dependence is similar to the
behavior of the thermal membrane height fluctuations
(Safran,1994): $\langle h^2(q)\rangle_{T}=k_{B}T/\kappa q^4$.The appearance of thermal-like correlations is not surprising,
since the driving force for the height fluctuations comes from the
thermal fluctuations of the membrane protein density: $\langle
f_{n}^2(q,\omega)\rangle\propto k_{B}T$. Thermal-like correlations
also appear for various choices of active membranes
(Manneville,20001;Ramaswamy,2000;Gov,2004). The
effective temperature we defined, has the following limiting forms
\begin{eqnarray}
D'\rightarrow 0\quad \frac{T_{eff}}{T}&\rightarrow&\frac{\kappa
v_{0}^2}{2D D' n_{0}\chi}=\frac{v_{0}}{2D |\overline{H}|}
\nonumber \\
D'\rightarrow \infty\quad
\frac{T_{eff}}{T}&\rightarrow&\frac{\kappa v_{0}^2}{2\sqrt{D D'^3}
n_{0}\chi}=\frac{1}{2}\sqrt{\frac{v_{0}n_{0}\chi}{D
\kappa|\overline{H}|^3}}
 \label{teff}
\end{eqnarray}
 The functional form of
$T_{eff}/T$ is very intuitive: the effective temperature increases with the pushing velocity of the actin $v_{0}$, and is inversely proportional to the diffusion coefficient of the membrane proteins, which smooths away the
density fluctuations.

In contrast, the density fluctuations are finite in the limit
$q\rightarrow0$. Note that for the free diffusion problem, we
recover the usual free diffusion: $\langle
n^2(q)\rangle=n_{0}q^2k_{B}TD/2w_{D}\sim n_{0}$. In our model we
find in the following limits
\begin{eqnarray}
D'\rightarrow 0\quad \langle n^2(0)\rangle&\rightarrow&\frac{
n_{0}k_{B}T}{\chi}\left(1-\sqrt{\frac{D'}{2D}}\right)
\nonumber \\
D'\rightarrow \infty\quad \langle n^2(0)\rangle&\rightarrow&\frac{
n_{0}k_{B}T}{\chi}\sqrt{\frac{D}{D'}}
 \label{n0q}
\end{eqnarray}
The first limit shows the approach to the bare membrane diffusion
in the absence of actin polymerization ($v_{0}\rightarrow0$). In
the second limit we find that the rapid formation of membrane
undulations, due to $v_{0}\rightarrow\infty$, effectively
localizes the membrane proteins and suppresses any long wavelength
density fluctuations.

In the limit $q\rightarrow\infty$ the density and height
fluctuations are given by
\begin{eqnarray}
\langle n^2(\infty)\rangle&=&\frac{\langle
f_{n}^2\rangle}{2D}=\frac{ n_{0}k_{B}T}{\chi}
\nonumber \\
\langle h^2(\infty)\rangle&=&\frac{16 \eta^2 v_{0}^2 k_{B}T}{\chi
n_0 \kappa^2} \frac{1}{q^6}
 \label{corrfuncinf}
\end{eqnarray}
The density fluctuations are finite, and approach the bare
membrane diffusion result (see first part of Eq.\ref{n0q}). The
height fluctuations decay in this limit in a non-thermal form,
 reminiscent of similar results for model active membranes
(Gov,2004).

In the case of negative spontaneous curvature of the membrane
proteins ($\overline{H}<0$), as we have already seen, no
instability occurs (fig.4b). For the density correlations we
 find that $\langle n^2(q)\rangle$ is approximately a
constant as a function of $q$ (Fig.4b), close to the limiting
value $\langle n^2(\infty)\rangle$ (Eq.\ref{corrfuncinf}) of free
diffusion. There is a region of reduced correlations, that
corresponds to the additional restoring curvature force, which now
acts to smooth away any density fluctuations. The density
correlations therefore dip for wavevectors $q<q_{w}$
(Eq.\ref{qw}), where there is wave-like propagation. The height
correlations show a cross-over from $1/q^4$ to $1/q^6$ decay
around $q_{w}$.

We now consider the correlations for the tension dominated case (or thin
lamellipodium edge). Here the membrane response is given by $\omega_{q,1},\omega_{q,t}\propto q$. The results in the low and high $q$ limits are 
\begin{eqnarray}
q\rightarrow 0\quad \langle
h^2(q)\rangle&\simeq&\frac{k_{B}T_{eff}}{\sigma q^{2}}\quad ,
\frac{T_{eff}}{T}=\frac{4 \pi v_{0}^2 \eta^2}{n_{0} \chi \sigma}
\nonumber \\
q\rightarrow \infty\quad \langle
h^2(q)\rangle&\simeq&\frac{k_{B}T_{eff}}{\kappa q^{4}}\quad
\nonumber \\ \frac{T_{eff}}{T}&=&\frac{v_{0}}{2D
|\overline{H}|}\frac{\sqrt{(2D'/D+1)+\sqrt{4D'/D+1}}-\sqrt{(2D'/D+1)-\sqrt{4D'/D+1}}}{\sqrt{4D'/D+1}}
 \nonumber \\ \label{tensionh}
\end{eqnarray}
The first limit has the same form as for thermal fluctuations in
the tension dominated regime, while the large $q$ limit has the
form of thermal fluctuations in a free and tension-less membrane, prompting us to express both in terms of an effective temperature.

One of the most striking results is that the height fluctuations increase with
increasing fluid viscosity $\eta$
(Eqs.\ref{divcorrfunc}-\ref{corrfuncinf},\ref{tensionh}).This behavior can be intuitively understood as arising from a height fluctuation that is driven at
a constant velocity $v_{0}$ over a time interval that is
proportional to the viscosity ($1/\omega_{q}\propto\eta$). This is
similar to the results of a previous analysis of active membranes
driven by ion pumps (Prost,1998). In both these cases, the active membrane proteins are allowed to have
density fluctuations, while they impose a given velocity on the
membrane. This is in contrast to the results of active membrane
proteins that produce a fluctuating force with zero average value
(shot-noise), where the height fluctuations are found to decrease
with increasing fluid viscosity (Gov,2004). Note that if the actin
imposes a force, $F_{actin}$, rather than a velocity $v$, on the
membrane, this introduces another factor of $1/q^2\eta^2$ to the
height correlations $\langle h^2(q)\rangle$ (see discussion
following Eq.\ref{eqmotion}), and the dependence on the
viscosity is subsequently modified.

The density and height fluctuations also increase when
the membrane diffusion coefficient is decreased
(Eqs.\ref{divcorrfunc},\ref{teff},\ref{tensionh}): $\langle
h^2(q)\rangle\propto1/D$. A similar result appeared in
(Prost,1998), but was limited there to the tension-less regime. We
find that this behavior also appears in the tension-dominated
case (Eq.\ref{tensionh}), which seems to be more realistic for
living cells (see next section). The origin of this behavior is
very intuitive; slower diffusion allows fluctuations in the
membrane protein density to survive longer, causing larger height undulations.

We now examine how the correlations depend on frequency. The power spectrum of the height
fluctuations as a function of frequency can be obtained by integrating Eq.\ref{corrfunc} over $q$. This gives us the temporal correlations $\langle
n^2(\omega)\rangle,\langle h^2(\omega)\rangle$. Since the integration
is not possible analytically, we do it numerically. We
plot the height correlation function in Fig.5, for the cell
membrane with the elastic parameters: $\kappa$ and $\sigma$. These
parameters are found empirically, by fitting to the observed
thermal fluctuations alone, i.e. when the actin polymerization is
blocked (Zidovska,2003) (see next section). For comparison we also
plot the correlation that arises purely from thermal fluctuations. This approaches the free membrane limit at high frequencies $\omega\rightarrow \infty$, where $\langle h^2(\omega)\rangle_{thermal}\rightarrow \omega^{-5/3}$
(Zilman,2002). At lower frequencies, in the tension dominated regime, the thermal behavior is: $\propto\omega^{-1}$.

The active fluctuations are found to have a $\langle
h^2(\omega)\rangle_{actin}\propto\omega^{-2}$ behavior at small
$\omega\rightarrow 0$, and $\langle
h^2(\omega)\rangle_{actin}\propto\omega^{-3}$ at large
$\omega\rightarrow \infty$. The cross-over occurs roughly where
the frequency of the membrane bending modes equals the
frequency of the effective diffusion of the membrane bumps. For the case when $\overline{H}<0$, the cross-over corresponds to the appearance of
propagating waves and occurs at the wavevector, $q_{w}$. The
frequency of the height fluctuations corresponding to this wavevector, is
given by $\omega_{h}$, shown as the vertical dashed-line in Fig.5.

Future experiments that probe the power spectrum of the height
fluctuations, should show a clear difference between the thermal
and active components. The actin-induced contribution to the
height fluctuations are predicted to be much more confined to low
frequencies than the thermal contribution.

\section{Discussion}

\subsection{Comparison with experiments}

Experimental observations of the actin-driven motion of cells and cell
membranes (Gerisch,2004;Vicker,2002;Giannone,2004) show both wave-like
propagations and finger-like filopodia. Since our model predicts that
 both behaviors are possible, depending on the sign of the spontaneous curvature, 
$\overline{H}$, of the membrane proteins, one possibility is that activators with both types of spontaneous curvature exist in vivo. There is also the
possibility that the spontaneous curvature of a membrane activator
is altered by a conformational change that is brought about by
phosphorylation or binding to another protein (or a number of
proteins), either already in the membrane or from the cytoplasm
(Small,2002). Thus the cell has many options, all of which it may use
to adjust the local concentration of the membrane proteins which
produce either uniform growth or filopodia
(Biyasheva,2004;Nozumi,2003). We predict that these proteins (or
protein complexes) have different spontaneous curvatures:
$\overline{H}<0$ for uniform growth and $\overline{H}>0$ for
filopodia. Indeed for filopodia growth the membrane proteins have
to form specific complexes which, in light of our model, must have
large spontaneous curvature (Gauthier-Campbell,2004;Wood,2002).
These complexes can then recruit crosslinking proteins such as fascin, which help form tight actin
bundles inside the growing filopodia
(Biyasheva,2004). Future extensions of our model may include the
coupling of dynamical changes in the spontaneous curvature,
$\overline{H}$, to the local densities of various proteins.

Another source of experimental corroboration for our overall
qualitative picture of the dynamic co-localization of activator
membrane proteins and actin polymerization in high (or low)
curvature regions of the leading edge, is provided by the
experiments of Nalbant {\it et al} (Nalbant,2004). The authors
visualized the dynamics of activated Cdc42 in living cells and
showed that there was a strong correlation between the most
recently formed protrusions and the level of active Cdc42 in its
vicinity, mostly concentrated at the tips of the protrusions. They
also show that the activator proteins are present only near the
base regions of filopodia, consistent with our assertion that the
activators form the initial seed for the filopodium by assembling
an actin rich bump, whereafter the newly recruited bundling agents
and normal barbed end polymerization of the actin filaments can
lead to the filopodia structure.

For positive spontaneous curvature of the membrane proteins
($\overline{H}>0$), we can estimate the critical wavevectors
(Eq.\ref{qc}), using typical values of the various parameters
(see sec.II). For the spontaneous curvature we use $\overline{H}\sim
(5-100nm)^{-1}$ (Girard,2005). For the flat membrane case we find:
$q_{c}^{-1}\simeq 1-10\mu$m. For the membrane edge case we find:
$q_{c,1}^{-1}\simeq d\cdot (dq_{c})$, where typically:
$d\sim0.5-1\mu$m. Our analysis predicts a specific wave-vector
($q_{c}$) which becomes unstable, so that the resulting filopodia
should have an average spacing given by the corresponding wavelength. This
length-scale appears to correlate well with the observed average
separation between neighboring filopodia (Oldenbourg,2000), of
$1.5-3\mu$m. Note that from Eq.\ref{qc}, increase in the membrane
tension causes an increase in the density of filopodia
(Parker,2002). Increased membrane tension was found to reduce the
velocity with which actin polymerization is pushing the membrane
(Raucher,2000), so that we expect not only more numerous but also smaller
filopodia under increased membrane tension.

We predict that the density and height fluctuations increase when
the membrane diffusion coefficient is decreased
(Eqs.\ref{divcorrfunc},\ref{teff},\ref{tensionh}). The membrane
diffusion coefficient may be changed by addition of various
chemical agents, such as changing the cholesterol content
(Vasanji,2004;Vrljic,2005). Note that changing the cholesterol
level may affect the activation of the membrane proteins
(Niggli,2004), which is a process that is not included in the
present work.

This prediction may explain the recently observed low membrane
diffusion (high micro-viscosity) at the leading edge of moving
cells (Vasanji,2004). The ratio of
the diffusion coefficients between the cell side and leading edge
is found to be: $D_{trail}/D_{lead}\sim3$. If we correlate the
r.m.s. height fluctuations from Eqs.(\ref{divcorrfunc},\ref{teff})
with the average rate of lamellipodial extension, we predict:
$V_{lead}/V_{side}\simeq\sqrt{\langle h^2(q)\rangle_{lead}/\langle
h^2(q)\rangle_{side}}\simeq\sqrt{D_{trail}/D_{lead}}\simeq\sqrt{3}$.
This is in good agreement with the measurement (Vasanji,2004). Our
model therefore provides a natural explanation for this otherwise paradoxical
observation: the membrane is stiffer (more viscous) in regions where motility
is increased. Presumably larger undulations in the shape of the
leading edge, help the cell overcome local friction barriers, and
results in faster overall motion (Ehrengruber,1996). Additionally,
the membrane undulations at the moving front can provide
localization points for the formation of adhesion complexes, which
are important in completing the cycle of cell motility
(Lavelin,2005;Zaidel-Bar,2003).

This result of our model may also explain the observed response of
endothelial cells' motion to shear flow
(Tardy,1997;Albuquerque,2003). In these experiments it was shown
that the cells move less quickly against the direction of the
flow, as compared to the perpendicular and parallel directions.
Concurrent with this motion, there is an increase in the fluidity
of the membrane in the front part of the cell, by as much as a
factor of 2 (Haidekker,2000;Butler,2001). According to our model
the amplitude of the active membrane fluctuations is therefore
reduced by a factor of $\sim\sqrt{2}$ compared to the rear of the
cell, which is in very good agreement with the measured drop in
the fluctuation amplitude in the presence of shear
(Dieterich,2000). This then results in the observed orientational
motility (Tardy,1997;Albuquerque,2003). Such a physical response
to shear flow may complement or trigger the biochemical changes
that take place in the presence of shear-flow (Zaidel-Bar,2005).

For negative spontaneous curvature of the membrane proteins
($\overline{H}<0$), we can estimate the critical wavevector (Eq.\ref{qw}) using the values of the parameters that appear
above. This gives us
$q_{w}^{-1}\simeq 1-10\mu$m, which is similar to what we obtained for $q_{c}$
(Eq.\ref{qc}). Thus both the instabilities and the wave-like motions
have the property that they occur only for membrane length-scales
larger than some critical length-scale, $\sim 1\mu$m. Indeed,
there are no long-lived actin structures smaller than this
length-scale, on the cell membrane (Gerisch,2004;Vicker,2002).

For the velocity of the propagating waves, we get from
Eq.\ref{vwave}: $v_{eff}\sim qD$, which results in velocities of
the order: $v_{eff}\sim 1-0.1\mu m/$sec for wavevectors: $q\sim
1(\mu m)^{-1}$. Actin waves with these velocities and wavelengths
are indeed observed on the surface of cells and lamellipodia
(Gerisch,2004;Vicker,2002;Giannone,2004). Note that in these
experiments the observed waves are on the bottom part of the cell,
where the membrane is largely flat next to the glass substrate.
Our analysis predicts that the actin waves correspond to small
undulations on the membrane surface (Fig.2b). Our interpretation
of these waves is therefore different from that given in (Vicker,2002),
where the surface waves are proposed to be sections through
three-dimensional spiral waves in the cell bulk. Recent
experiments seem to confirm our interpretation since they suggest that the actin structures are
largely confined to the cell membrane (Gerisch,2004), and that the
travelling-wave propagation is related to actin polymerization
around a high density Arp2/3 complex (Bretschneider,2004). The
formation and decay of these density fluctuations occurs on a timescale,
$t_{fluct}\sim2-3$sec, which is in agreement with membrane diffusion times over
the lateral size of these formations ($\sim1/2\mu$m). The slower
decay as compared to the formation, $\sim3.5$sec vs. $\sim2$sec
respectively, may be due to the extra distance to diffuse out of
the bump (Fig.2b): $t_{fluct}\cdot v_{0}\sim0.1\mu$m.
Alternatively the diffusion coefficient may decrease due to the
dense actin gel formation.

Recently the static height correlations, $\langle h^2(q)\rangle$,
were measured on living cells (Zidovska,2003).  The mean-square
height undulations of the active cell are found to be $\sim8$
times larger than for the inactive cell (Zidovska,2003). The
correlations agree with the tension-dominated behavior given in
Eq.\ref{tensionh} (Fig.6), if we use the same parameters as before and
take the surface tension to be $\sigma\sim0.5\times10^{-8}$J/m$^2$. In particular, cells
that lack the actin-polymerization motility, display much smaller
fluctuations (Zidovska,2003) presumably of thermal origin. Indeed
these fluctuations are well described by confined thermal
correlations of the form (Gov,2003): $\langle h^2(q)\rangle\propto
k_{B}T/(\kappa q^4 +\sigma q^2 +\gamma)$, with
$\sigma\sim3.7\times10^{-8}$J/m$^2$ and
$\gamma\sim2.6\times10^{5}$J/m$^4$ (Fig.6). The equivalent
confinement distance: $d_{T}\simeq
k_{B}T/8\sqrt{\gamma\kappa}\sim70$nm, is consistent with the
average separation between the fibers of the actin mesh, which
underlies the membrane. In a normal cell, when actin polymerization is driving the membrane
fluctuations, there is no meaning to any
membrane confinement. It is then possible to fit the active cell
data to the thermal fluctuations of an unconfined membrane, with
an "effective temperature" of $T_{eff}/T\sim8$ (Fig.6). This
approach though does not give us any information about the nature
of the active fluctuations. From the $q\rightarrow0$ limit of
Eq.\ref{tensionh} we get an effective temperature of this
magnitude if we take: $\sigma$ to be smaller than the value given
by the fit to the thermal fluctuations, by a factor of $\sim8$,
$n_0\sim(300nm)^{-2}$, and an effective viscosity
$\eta\sim100\eta_{water}$. These parameters are within reasonable
limits for a cell, but an exact comparison with the data awaits
independent determination of these parameters.

Other experimental data (Agero,2003) indicate that in a living
cell there is an exponential component to the probability
distribution function of the height fluctuations, while in a
cell that has its actin polymerization blocked, only the usual
Gaussian term due to thermal membrane fluctuations remains. We can
assume that the usual thermal height fluctuations add incoherently
to the active fluctuations, which according to our model arise
from the fluctuations in the density of the membrane proteins, $n$.
For simplicity let us describe the case where $\overline{H}=0$, so
that the density of the membrane proteins is decoupled from the
membrane height fluctuations (Eq.\ref{diffusion}). In the 
limit of small density fluctuations, the corresponding
actin polymerization induced height fluctuations are given by
Eq.\ref{eqmotion}
\begin{equation}
\delta h\simeq A\delta n dt\sim \frac{A L^2}{n_{0}D} \delta
n^2\label{deltah}
\end{equation}
where the diffusion occurs in a membrane patch of size $L$ over a
time: $dt\simeq (L^2/D)(\delta n/n_0)$. This relation states that
larger density fluctuations (larger $\delta n/n_0$) take longer to diffuse
away. The probability distribution function of the density
fluctuations $\delta n/n_0$ is a thermal Gaussian
(Manneville,2001;Ramaswamy,2000), so that combined with
Eq.\ref{deltah} we get
\begin{eqnarray}
P(\delta n)&\simeq&e^{(-\chi \delta n^2/n_{0}^2k_{B}T)}
\nonumber \\
\Rightarrow P(\delta h)&\simeq&e^{(-\chi D \delta h/v_{0} L^2
k_{B}T)} \label{prob}
\end{eqnarray}
Thus the probability distribution does indeed have an exponential component.
The larger the driving velocity $v_0$, the more enhanced the
distribution for larger height fluctuations. On the other hand
large values of the diffusion coefficient $D$ or membrane protein compression
energy $\chi$, narrow the distribution by causing density
fluctuations to have shorter life-times and amplitudes. A
quantitative comparison with the observations awaits future
experiments.

Recent data (Neto,2005) indicates that the velocity of
actin induced membrane ruffles, is strongly temperature dependent.
According to our model this velocity is proportional to the
membrane diffusion coefficient (Eq.\ref{vwave},\ref{vwave1}):
$v_{eff}\propto Dq$, where $q$ is the inverse of the typical
length-scale of these ruffles (usually $\sim1-2\mu$m (Neto,2005)).
Since the diffusion coefficient is inversely proportional to the
membrane viscosity (Almeida,1995), we expect it to vanish at the
liquid-gel transition temperature $T_{m}$, where the viscosity
diverges (Dimova,2000): $D(T)\propto|T-T_{m}|^{1.4}$. Using
typical values for $T_m=20^oC$, and $q=0.67 \mu m^{-1}$, we fit
the overall scale of $D(T)$ (inset of Fig.7), such that the
resulting velocity $v_{eff}$ agrees with the observation (Fig.7a).
Furthermore, the observed decay time of smaller height
fluctuations, also behaves as: $\tau_{decay}\propto1/Dq^2$, with a
larger wavevector (smaller wavelength) of $q=1.15 \mu m^{-1}$
(Fig.7b).

Finally, the mean-square membrane curvature was observed not to
depend on the temperature (Neto,2005). The mean-square curvature:
$\langle H^2\rangle=\int q^4 h_q^2 d^2q$, is dominated by
$q\rightarrow \infty$ modes, while the mean-square amplitude of
height fluctuations $\langle h^2\rangle$ is dominated by the
$q\rightarrow 0$ modes. From our model we predict that for the
tensionless membrane, the amplitude of the $q\rightarrow 0$ modes
does depend on the diffusion $D$ (Eq.\ref{teff}): $\langle
h^2\rangle\propto1/D(T)$, while the amplitude of the $q\rightarrow
\infty$ modes does not (Eq.\ref{corrfuncinf}). This difference
could explain the independence of the observed r.m.s. curvature on
the temperature, except for the very weak $\sqrt{\langle H^2\rangle} \propto \sqrt{k_{B}T}$ which amounts to $\sim 2\%$ over the observed temperature range (Neto,2005).

Detailed comparisons between our model (namely $\langle
h^2(q)\rangle$,$\langle h^2(\omega)\rangle$) and the experimental
data (Gerisch,2004;Bretschneider,2004), awaits more quantitative
analysis of the spatial and temporal shape fluctuations in living
cells (Zidovska,2003;Agero,2003;Neto,2005).

So far we have discussed the membrane dynamics at the leading edge which is what we are modeling. However, our model can also give us insights into phenomena that occur at the cellular scale.
An example of complicated, oscillatory dynamics of the bulk actin
gel, is described in (Giannone,2004). The authors find a periodic interruption ($\sim 20$s) in the forward motion of lamellipodia that had no filopodia. One possibility for the
mechanism is shown schematically in (Fig.8). The activating membrane proteins are
initially concentrated at the lamellipodium edge, and since there
are no filopodia we can take the spontaneous curvature to be small
or negative. As the forward motion persists the leading edge thins
such that the local curvature is too high for these proteins and
they prefer to move to the less curved membrane on the upper
surface. This causes a backward propagating wave of actin
polymerization, which proceeds until the back edge of the
lamellipodium. This explains why the contractions occur every
$t_{cont}\sim L_{lam}/v\sim10-30$sec, where $L_{lam}\sim2\mu$m is
the thickness of the lamellipodium (Giannone,2004) and
$v\sim0.1\mu$m/sec is of the order of the calculated propagation
velocity $v_{eff}$ (Section IV) (Fig.8). The membrane dynamics we
considered in the present model are therefore coupled in the cell
to the dynamical variations affecting the entire actin layer, and
this coupling remains to be described.

\subsection{Predictions}
Our model allows us to make testable and quantitative predictions.
For example:
\begin{itemize}
    \item Changing the fluid viscosity and membrane tension will shift the average
    density of filopodia (Eq.\ref{qc}).
    \item The velocity of propagation of actin density fluctuations on the cell membrane,
    is predicted to be linear in the diffusion coefficient of membrane
    proteins (Eq.\ref{vwave}).
    \item Similarly, the amplitude of density and height fluctuations are predicted to increase when
the membrane diffusion coefficient is decreased (Vasanji,2004)
(Eqs.\ref{divcorrfunc},\ref{teff},\ref{tensionh}). Note that the
density of filopodia, given by $q_{c}$ (\ref{qc}), is independent
on $D$.
    \item In the tension-dominated regime, which seems to be applicable to most
    cells, the long wave-length height fluctuations increase with increasing of the
    fluid viscosity (Eq.\ref{tensionh}). The same behavior is
    found also in the tension-less case (Eq.\ref{teff}).
    \item Our prediction that the membrane proteins that initiate
    filopodia (Nozumi,2003) have a high spontaneous curvature, has
    to be tested. Incorporation of these proteins into synthetic
    vesicles and observing the resulting shape transformations
    could determine this parameter.
\end{itemize}

Some of these manipulations are possible in living cells, while
others are better tested in synthetic systems (Vasanji,2004).

\section{Conclusion}

The dramatic membrane dynamics that occur at the surface of stimulated cells is a consequence, not only of the actin polymerization dynamics, but also of the interplay between the dynamics of the membrane itself and that of the activators that reside on it. Keeping this in mind,
we presented a simple model that treats the dynamics of a membrane under the action of actin polymerization forces that depend on the local density of freely diffusing activators on the membrane. We took into account the thermal density fluctuations and the spontaneous curvature associated with the activators and showed that, depending on the spontaneous membrane curvature associated with the activators, the resulting membrane motion can be wave-like, corresponding to membrane ruffling and actin-waves, or unstable, indicating the tendency of filopodia to form. Thus our simple model system managed to capture the wide range of complex dynamics observed at the leading edges of motile cells both qualitatively and quantitatively indicating that the essential physics had been retained. Our model not only provides detailed estimates of the morphology and dynamics of the membrane structures, but also provides quantitative explanations for a variety of related experimental observations. These include the puzzling increase in membrane micro-viscosity at the leading edge of migrating cells, the appearance of an exponential contribution to the probability distribution of membrane height fluctuations and the temperature dependence of the membrane ruffle velocity among others. Thus, our model offers a simple framework with which to analyze and understand  experimental data and make quantitative predictions to be tested by future experiments.

We should, however, keep in mind that cell motility involves
many processes which we did not take into account in our model,
such as adhesion, formation of stress fibers and the action of
molecular motors. Even within the context of our model, the dynamics of the actin-driven cell motility is largely assumed to be controlled by dynamics of proteins on the cell membrane, which fails to capture the link  between the dynamics of these membrane proteins and the bulk dynamics of the actin gel, occurring behind the moving front (Plastino,2004). Integrating all these components into a holistic picture remains a challenge. We should therefore view our model as
representing the physical dynamics of the membrane-actin system,
which trigger the formation of patterns in the membrane
morphology, that are part of the overall motility mechanism.
Our model therefore provides answers to one part of the overall problem of cell motility and should be useful for any integrated approach to cellular motility.

\begin{acknowledgments}
N.G. thanks BSF grant number 183-2002, EU SoftComp NoE grant and
the Robert Rees Fund for Applied Research, for their support. A.G.
would like to acknowledge support from MRL Program of the NSF
under Award number DMR00-80034 and NSF Grant number DMR02-037555.
N.G. is also thankful for the kind hospitality of Phillip Pincus
at the MRL, University of California Santa-Barbara, where this
research was initiated.
\end{acknowledgments}

\section{References}

Agero, U., C.H. Monken and C. Ropert and R.T. Gazzinelli and
O.N. Mesquita.2003. Cell surface fluctuations studied with
defocusing microscopy. Phys. Rev. E. 67(5 Pt 1):051904.

Albuquerque, M.L., and A.S. Flozak.2003. Lamellipodial motility in
wounded endothelial cells exposed to physiologic flow is
associated with different patterns of beta1-integrin and vinculin
localization. J. Cell Physiol. 195(1):50-60.

Almeida, P.F.F., and W.L.C. Vaz.1995. Handbook of Biological
Physics. Volume 1, edited by R. Lipowsky and E. Sackmann, Elsevier
Science B.V. 305.

Bailly, M., F. Macaluso, M. Cammer, A. Chan, J.E.
Segall and J.S. Condeelis.1999. Relationship between Arp2/3
complex and the barbed ends of actin filaments at the leading edge
of carcinoma cells after epidermal growth factor stimulation. J.
Cell Biol. 145(2):331-345.

Biyasheva, A., T. Svitkina, P. Kunda, B. Baum and G.
Borisy.2004. Cascade pathway of filopodia formation downstream of
SCAR. J. Cell. Sci. 117(Pt 6):837-848.

Blanchoin, L., K.J. Amann, H.N. Higgs, J.B. Marchand,
D.A. Kaiser and T.D. Pollard.2000. Direct observation of dendritic
actin filament networks nucleated by Arp2/3 complex and WASP/Scar
proteins. Nature. 2000 404(6781):1007-1011.

Bottino, D., A. Mogilner, T. Roberts, M. Stewart and G.
Oster.2002. How nematode sperm crawl. J. Cell Sci. 115(Pt
2):367-384.

Bretschneider, T., S. Diez, K. Anderson, J. Heuser,
M. Clarke, A. Muller-Taubenberger, J. Kohler and G.
Gerisch.2004. Dynamic actin patterns and Arp2/3 assembly at the
substrate-attached surface of motile cells. Curr. Biol.
14(1):1-10.

Butler, P.J., G. Norwich, S. Weinbaum and S. Chien.2001.
Shear stress induces a time- and position-dependent increase in
endothelial cell membrane fluidity. Am. J. Physiol. Cell Physiol.
280(4):C962-969.

Carlier, M.F., C. Le Clainche, S. Wiesner and D.
Pantaloni.2003. Actin-based motility: from molecules to movement.
Bioessays. 25(4):336-345.

Carlsson, A.E.2003. Growth velocities of branched actin networks.
Biophys J. 84(5):2907-2918.

Chen, H.Y.2004. Internal states of active inclusions and the
dynamics of an active membrane. Phys. Rev. Lett. 92(16):168101.

Dieterich, P., M. Odenthal-Schnittler, C. Mrowietz,  M.
Kramer, L. Sasse, H. Oberleithner and H.J. Schnittler.2000.
Quantitative morphodynamics of endothelial cells within confluent
cultures in response to fluid shear stress. Biophys. J.
79(3):1285-1297.

Dimova, R., B. Pouligny, and C. Dietrich.2000. Pretransitional
Effects in Dimyristoylphosphatidylcholine Vesicle Membranes:
Optical Dynamometry Study. Biophys. J. 79(1):340–356.

Ehrengruber, M.U., D.A. Deranleau and T.D. Coates.1996. Shape
oscillations of human neutrophil leukocytes: characterization and
relationship to cell motility. J. Exp. Biol. 199(Pt 4):741-747.

Ford, M.G.J., I.G. Mills, B.J. Peter, Y. Vallis, G.J.K. Praefcke,
P.R. Evans and H.T. McMahon.2002. Curvature of clathrin-coated
pits driven by epsin. Nature. 419:361-366.

Gauthier-Campbell, C., D.S. Bredt, T.H. Murphy and Ael-D.
El-Husseini.2004. Regulation of dendritic branching and filopodia
formation in hippocampal neurons by specific acylated protein
motifs. Mol. Biol. Cell. 15(5):2205-2217.

Gerbal, F., P. Chaikin, Y. Rabin and J. Prost.2000. An
elastic analysis of Listeria monocytogenes propulsion. Biophys. J.
79(5):2259-2275.

Gerisch, G., T. Bretschneider, A. Muller-Taubenberger, E. Simmeth,
M. Ecke, S. Diez and K. Anderson.2004. Mobile actin clusters and
travelling waves in cells recovering from actin depolymerization.
Biophys. J. 87(5):3493-3503.

Giannone, G., B.J. Dubin-Thaler, H.G. Dobereiner, N.
Kieffer, A.R. Bresnick and M.P. Sheetz.2004. Periodic
lamellipodial contractions correlate with rearward actin waves.
Cell. 116(3):431-443.

Girard, P., J. Prost and P. Bassereau.2005. Active Pumping
Effects in Vesicle Fluctuations. Phys. Rev. Lett., in press.

Gov, N., A.G. Zilman and S. Safran.2003. Cytoskeleton
confinement and tension of red blood cell membranes. Phys. Rev.
Lett. 90(22):228101.

Gov, N.2004. Membrane undulations driven by force fluctuations of
active proteins. Phys. Rev. Lett. 93(26):268104.

Grimm, H.P., A.B. Verkhovsky, A. Mogilner and J.J.
Meister.2003. Analysis of actin dynamics at the leading edge of
crawling cells: implications for the shape of keratocyte
lamellipodia. Eur. Biophys. J. 32(6):563-577.

Gungabissoon, R.A., and J.R. Bamburg.2003. Regulation of growth
cone actin dynamics by ADF/cofilin. J. Histochem. Cytochem.
51(4):411-420.

Habermann, B.2004. The BAR-domain family of proteins: a case of
bending and binding? EMBO Rep. 5(3):250-255.

Haidekker, M.A., N. L'Heureux and J.A. Frangos.2000. Fluid
shear stress increases membrane fluidity in endothelial cells: a
study with DCVJ fluorescence. Am. J. Physiol. Heart Circ. Physiol.
278(4):H1401-H1406.

Higgs, H.N., and T.D. Pollard.2001. Regulation of actin filament
network formation through ARP2/3 complex: activation by a diverse
array of proteins. Annu. Rev. Biochem. 70:649-676.

Lavelin, I., and B. Geiger.2005. Characterization of a Novel
GTPase-activating Protein Associated with Focal Adhesions and the
Actin Cytoskeleton. J. Biol. Chem. 280(8):7178-7185.

Manneville, J.B., P. Bassereau, S. Ramaswamy and J. Prost.
2001. Active membrane fluctuations studied by micropipet
aspiration. Phys. Rev. E. 64(2 Pt 1):021908.

Martin, P., A.J. Hudspeth and F. J\"{u}licher. 2001. Comparison of
a hair bundle's spontaneous oscillations with its response to
mechanical stimulation reveals the underlying active process. PNAS
98 (25):14380–14385.

Mogilner, A., and G. Oster.1996. The physics of lamellipodial
protrusion. Eur. Biophys. J. 25(1):47-53.

Nalbant, P., L. Hodgson, V. Kraynov, A. Toutchkine and
K.M. Hahn.2004. Activation of endogenous Cdc42 visualized in
living cells. Science. 10;305(5690):1615-1619.

Neto, J.C. , U. Ageroa, D.C.P. Oliveirac, R.T.
Gazzinelli and O.N. Mesquita.2005. Real-time measurements of
membrane surface dynamics on macrophages and the phagocytosis of
Leishmania parasites. Exp. Cell Res. 303:207–217.

Niggli, V., A.V. Meszaros, C. Oppliger and S. Tornay.2004.
Impact of cholesterol depletion on shape changes, actin
reorganization, and signal transduction in neutrophil-like HL-60
cells. Exp. Cell Res. 296(2):358-368.

Nozumi, M., H. Nakagawa, H. Miki, T. Takenawa and S.
Miyamoto.2003. Differential localization of WAVE isoforms in
filopodia and lamellipodia of the neuronal growth cone. J. Cell
Sci. 116(Pt 2):239-246.

Oldenbourg, R., K. Katoh and G. Danuser.2000. Mechanism of
lateral movement of filopodia and radial actin bundles across
neuronal growth cones. Biophys. J. 78(3):1176-1182.

Parker, K.K., A.L. Brock, C. Brangwynne, R.J. Mannix,
N. Wang, E. Ostuni, N.A. Geisse, J.C. Adams, G.M.
Whitesides and D.E. Ingber.2002. Directional control of
lamellipodia extension by constraining cell shape and orienting
cell tractional forces. FASEB J. 16(10):1195-1204.

Plastino, J., I. Lelidis, J. Prost and C. Sykes.2004. The
effect of diffusion, depolymerization and nucleation promoting
factors on actin gel growth. Eur. Biophys. J. 33(4):310-320.

Pollard, T.D., and G.G. Borisy.2003. Cellular motility driven by
assembly and disassembly of actin filaments. Cell. 112(4):453-465.

Prost, J., J.-B. Manneville and R. Bruinsma.1998.
Fluctuation-magnification of non-equilibrium membranes near a
wall. Eur. Phys. J. B 1:465-480.

Ramaswamy, S., J. Toner and J. Prost.2000. Nonequilibrium
fluctuations, travelling waves, and instabilities in active
membranes. Phys. Rev. Lett. 84(15):3494-3497.

Raucher, D., and M.P. Sheetz.2000. Cell spreading and
lamellipodial extension rate is regulated by membrane tension. J.
Cell Biol. 148(1):127-136.

Safran, S.A.1994. Statistical Thermodynamics of Surfaces,
Interfaces and Membranes. Frontiers in physics v.90.
Addison-Wesley Publishing Company.

Sankararaman, S., G.I. Menon and P.B. Kumar.2004.
Self-organized pattern formation in motor-microtubule mixtures.
Phys. Rev. E 70(3 Pt 1):031905.

Small, J.V., T. Stradal, E. Vignal and K. Rottner.2002. The
lamellipodium: where motility begins. Trends Cell Biol. 12 (3):
112-120

Stephanou, A., M.A. Chaplain and P. Tracqui.2004. A
mathematical model for the dynamics of large membrane deformations
of isolated fibroblasts. Bull. Math. Biol. 66(5):1119-1154.

Tardy, Y., N. Resnick, T. Nagel, M.A. Gimbrone Jr and
C.F. Dewey Jr.1997. Shear stress gradients remodel endothelial
monolayers in vitro via a cell proliferation-migration-loss cycle.
Arterioscler. Thromb. Vasc. Biol. 17(11):3102-3206.

Vasanji, A., P.K. Ghosh, L.M. Graham, S.J. Eppell and
P.L. Fox.2004. Polarization of plasma membrane microviscosity
during endothelial cell migration. Dev. Cell. 6(1):29-41.

Vicker, M.G.2002. Eukaryotic cell locomotion depends on the
propagation of self-organized reaction-diffusion waves and
oscillations of actin filament assembly. Exp. Cell Res.
275(1):54-66.

Vrljic, M., S.Y. Nishimura, W.E. Moerner and H.M.
McConnell.2005. Cholesterol Depletion Suppresses the Translational
Diffusion of Class II Major Histocompatibility Complex Proteins in
the Plasma Membrane. Biophys. J. 88(1):334-347.

Wood, W., and P. Martin.2002. Structures in focus-filopodia. Int.
J. Biochem. Cell Biol. 34(7):726-730.

Zaidel-Bar, R., C. Ballestrem, Z. Kam and B. Geiger.2003.
Early molecular events in the assembly of matrix adhesions at the
leading edge of migrating cells. J. Cell Sci. 116(Pt
22):4605-4613.

Zaidel-Bar, R., Z. Kam and B. Geiger.2005. EMBO J. submitted
for publication.

Zidovska, A.2003. Diplomarbeit von Alexandra Zidovska:
"Micromechanical properties of the cell envelope and membrane
protrusions of Macrophages". Technische Universit\"{a}t
M\"{u}nchen Fakult\"{a}t f\"{u}r Physik Lehrstuhl f\"{u}r
Biophysik, Prof. Dr. Erich Sackmann.

Zilman, A.G., and R. Granek.2002. Membrane dynamics and structure
factor. Chemical Physics. 284 (1-2):195-204.

\newpage

\section{Figure Captions}

Fig.1: Schematic picture of the model. (a) The Arp activating
membrane proteins are symbolized by the black squares, diffusing
in the flat membrane. Where they have a high density the actin
polymerization is more extensive (dashed regions) and so is the
velocity of the membrane (normal arrows). (b) In the case where
the polymerization is confined to a thin leading edge, the local
high curvature changes the response to $\omega_{q,1}$.

Fig.2: Schematic picture of the two behaviors depending on the
spontaneous curvature of the membrane proteins. (a)
$\overline{H}>0$: fluctuations in the density of the Arp
activating membrane proteins grow unstable when the proteins
aggregate into the high curvature "filopodia". (b)
$\overline{H}<0$: wave-like propagation due to the restoring force
of the curvature, breaking up high density fluctuations (dashed
arrows).

Fig.3: Calculated response frequencies of the membrane protein
density $\omega_{n}$ and membrane height undulations $\omega_{h}$
(using $\omega_{q}$ of a flat membrane, $\kappa=10k_{B}T$ and
$v_0=1\mu$m/sec). The bare diffusion rate $\omega_{D}$ and bare
membrane response $\omega_{q}$ are shown by the dotted and
dot-dash line respectively. (a) $\overline{H}=(10nm)^{-1}$:
$\omega_{n}$-solid line, $\omega_{h}$-dashed line. The critical
wavevector $q_{c}$ (Eq.\ref{qc}) below which the density
fluctuations are unstable is indicated by the vertical dotted
line. (b) $\overline{H}=-(3nm)^{-1}$: The imaginary parts are
given by the solid lines while the real parts are given by the
dashed lines. The critical wavevector $q_{w}$ (Eq.\ref{qw}) below
which the wave-like fluctuations occur is indicated by the
vertical dotted line. In the inset we show that the imaginary part
can be smaller than the real part for small enough $q$.

Fig.4: Calculated static density and membrane height correlation
functions: $\langle n^2(q)\rangle$-solid line,$\langle
h^2(q)\rangle$-dashed line. (a) $\overline{H}=(10nm)^{-1}$: The
correlations diverge at the critical wavevector $q_{c}$
(Eq.\ref{qc}), indicated by the vertical dotted line. The limiting
values $\langle n^2(0)\rangle$ (Eq.\ref{n0q}) and $\langle
n^2(\infty)\rangle$ (Eq.\ref{corrfuncinf}) are shown by the
horizontal dotted and dash-dot line respectively. (b)
$\overline{H}=-(3nm)^{-1}$: The density correlations dip for
wavevectors $q<q_{w}$ (Eq.\ref{qw}), indicated by the vertical
dotted line. The height correlations show a monotonous decay,
having a cross-over from $1/q^4$ to $1/q^6$ behavior around
$q_{w}$. The limiting value $\langle n^2(\infty)\rangle$
(Eq.\ref{corrfuncinf}) is shown by the horizontal dash-dot line.

Fig.5: Calculated height fluctuations as a function of the
frequency $\omega$. The dash-dot line is the thermal fluctuations,
that approach $\omega^{-5/3}$ at large frequencies (asymptotic
dashed-line). The actin-induced fluctuations are given by the
solid and dotted lines, for $\overline{H}>0$ and $\overline{H}<0$
respectively. The asymptotic behavior is given by the dashed
straight lines: $\omega^{-2}$ and $\omega^{-3}$ in the limit of
small and large frequencies respectively. The vertical dashed line
represents the frequency of the cross-over, roughly given by
$\omega_{h}(q_{w})$.

Fig.6: Calculated static height correlation function $\langle
h^2(q)\rangle$ in the tension-dominated regime (Eq.\ref{tensionh})
(dashed line), compared with the data (Zidovska,2003) for normal
cell (stars) and inactivated cell (squares). The solid line and
dash-dot lines show the behavior for the thermal height
fluctuations in an unconfined and confined membrane respectively.
The lower panel shows the two cases: (a) Inactivated cell with
confined thermal membrane fluctuations, and (b) actin-induced
fluctuations in the normal cell.

Fig.7:(a) Calculated velocity of actin-induced membrane ruffles:
$v_{eff}\propto D(T)$ (solid line), where $D(T)$ is shown in the
inset. The experimental data (circles) is from (Neto,2005). (b)
Calculated decay time of actin-induced membrane fluctuations:
$\tau_{decay}\propto 1/D(T)$ (solid line), compared to the
experimental data (squares) (Neto,2005).

Fig.8: Schematic description of the periodic contractions found in
lamellipodia growth (Giannone,2004). (a) The membrane proteins
initially localized at the lamellipodia's edge (filled circles)
produce a forward pushing actin network (shaded ellipse). (b,c) As
the edge thins they are pushed towards less curved regions at the
top of the lamellipodia, and then propagate until the back edge.
(d) The forward edge is again thicker now, so these membrane
proteins can localize there and forward motion resumes.

\newpage

\section{Figures}

\begin{figure}
\centerline{\ \epsfysize 8cm \epsfbox{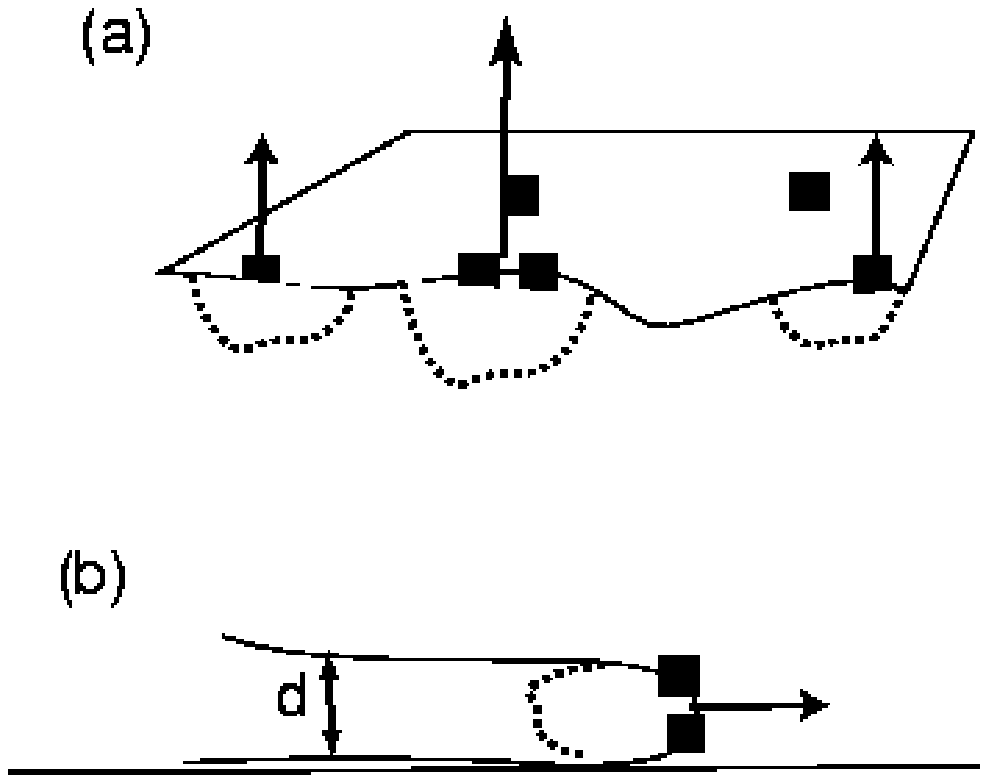}} \caption{}
\end{figure}

\begin{figure}
\centerline{\ \epsfysize 15cm \epsfbox{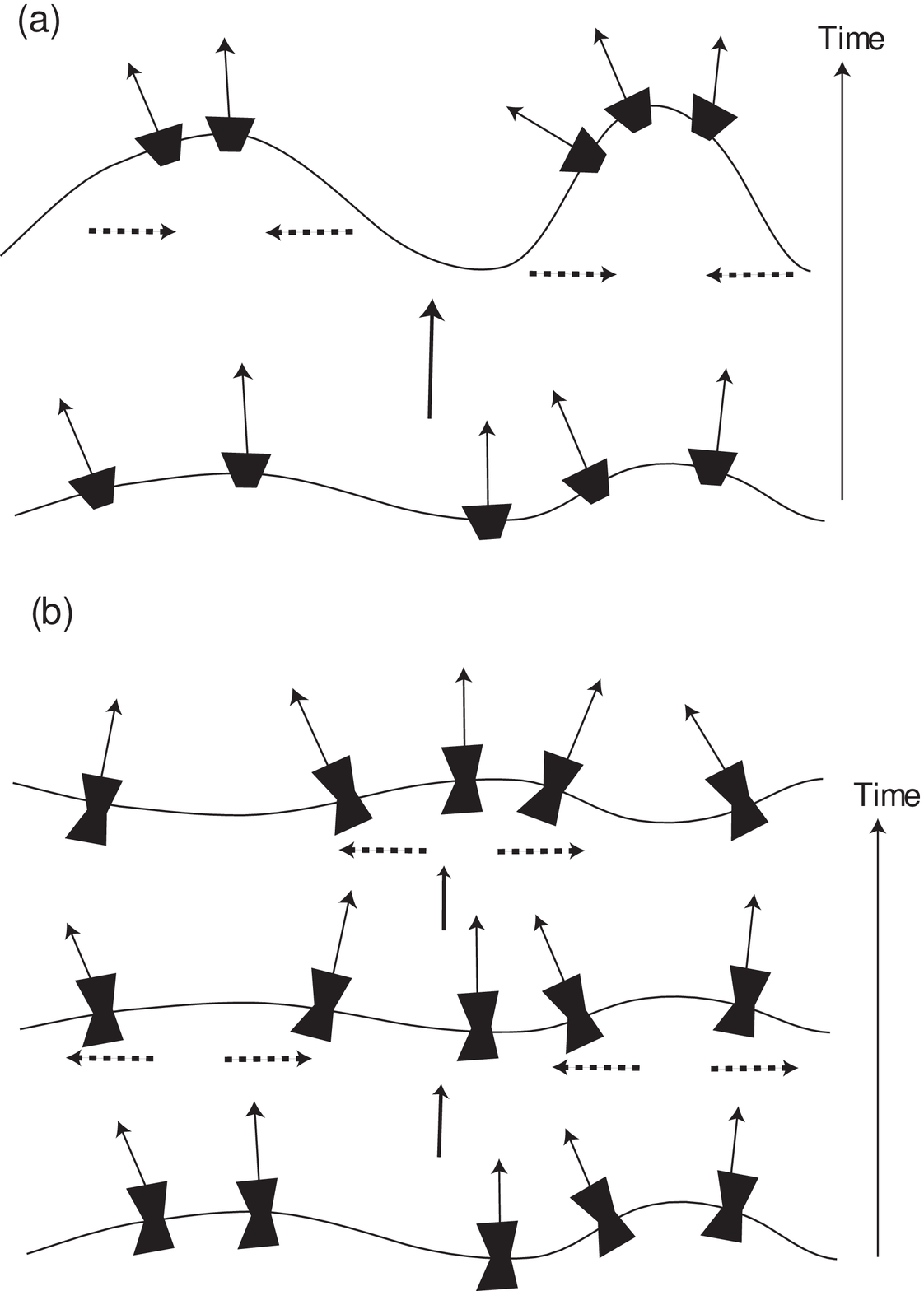}} \caption{}
\end{figure}

\begin{figure}
\centerline{\ \epsfysize 17cm \epsfbox{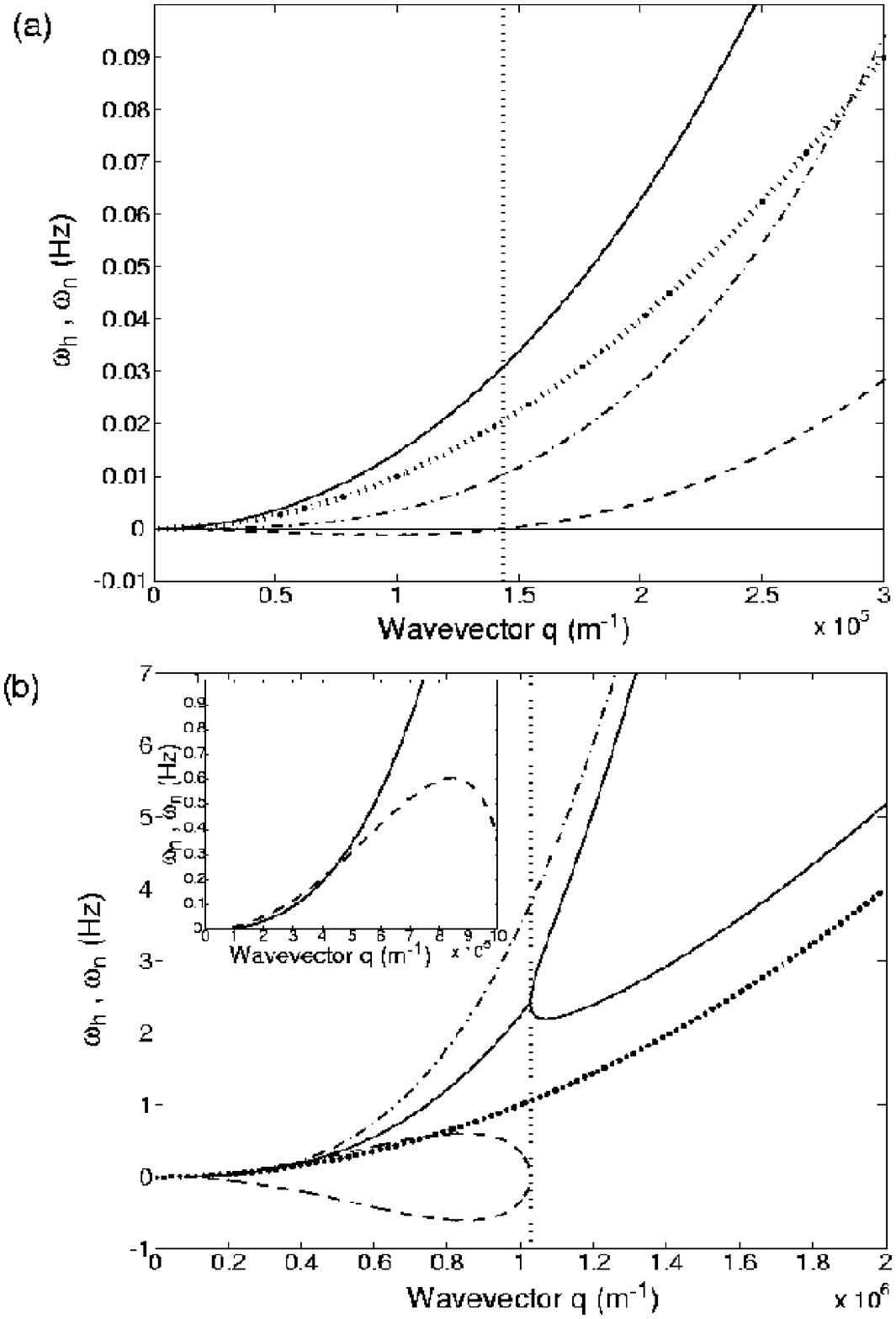}} \caption{}
\end{figure}

\begin{figure}
\centerline{\ \epsfysize 17cm \epsfbox{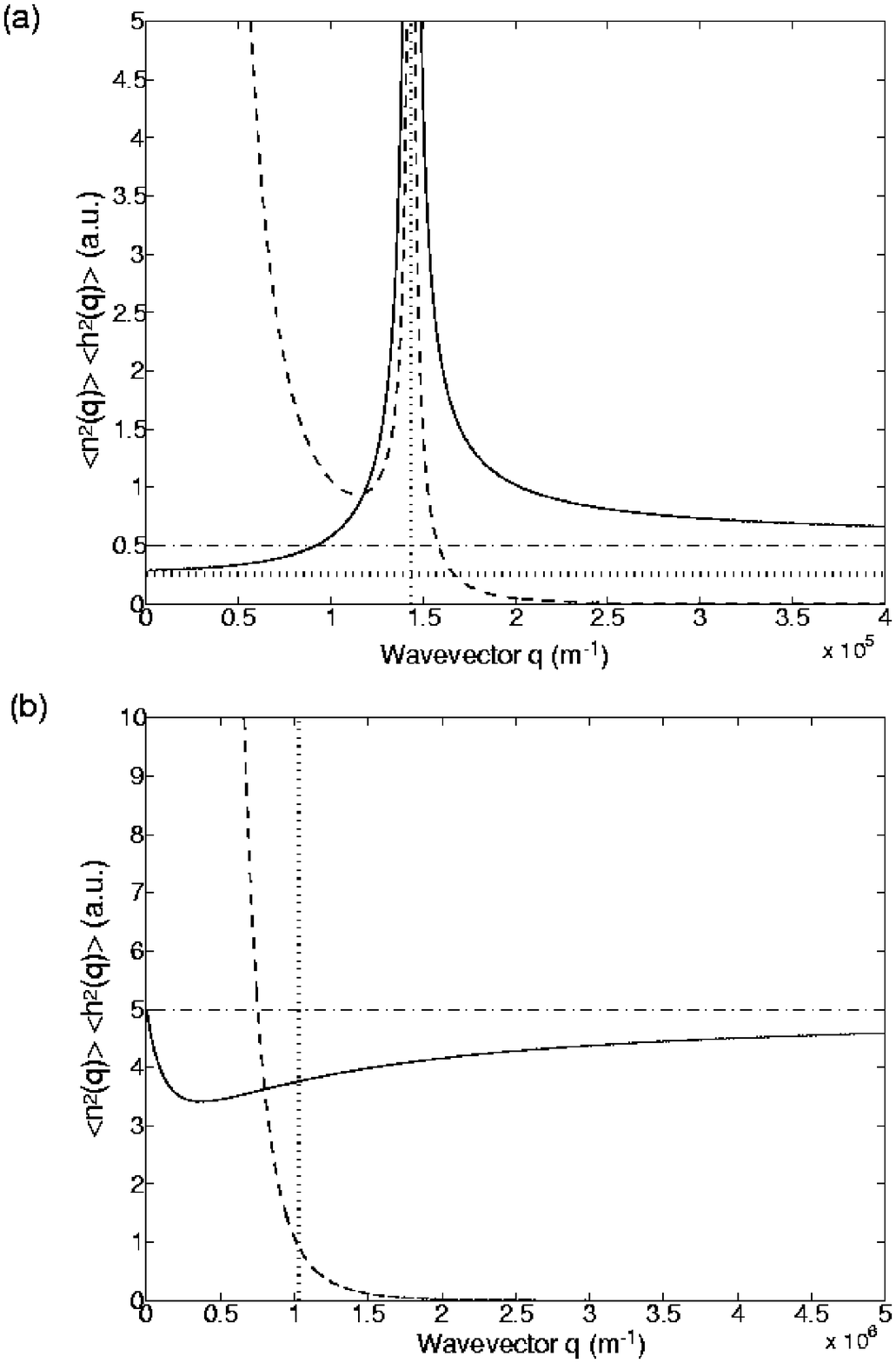}} \caption{}
\end{figure}

\begin{figure}
\centerline{\ \epsfysize 12cm \epsfbox{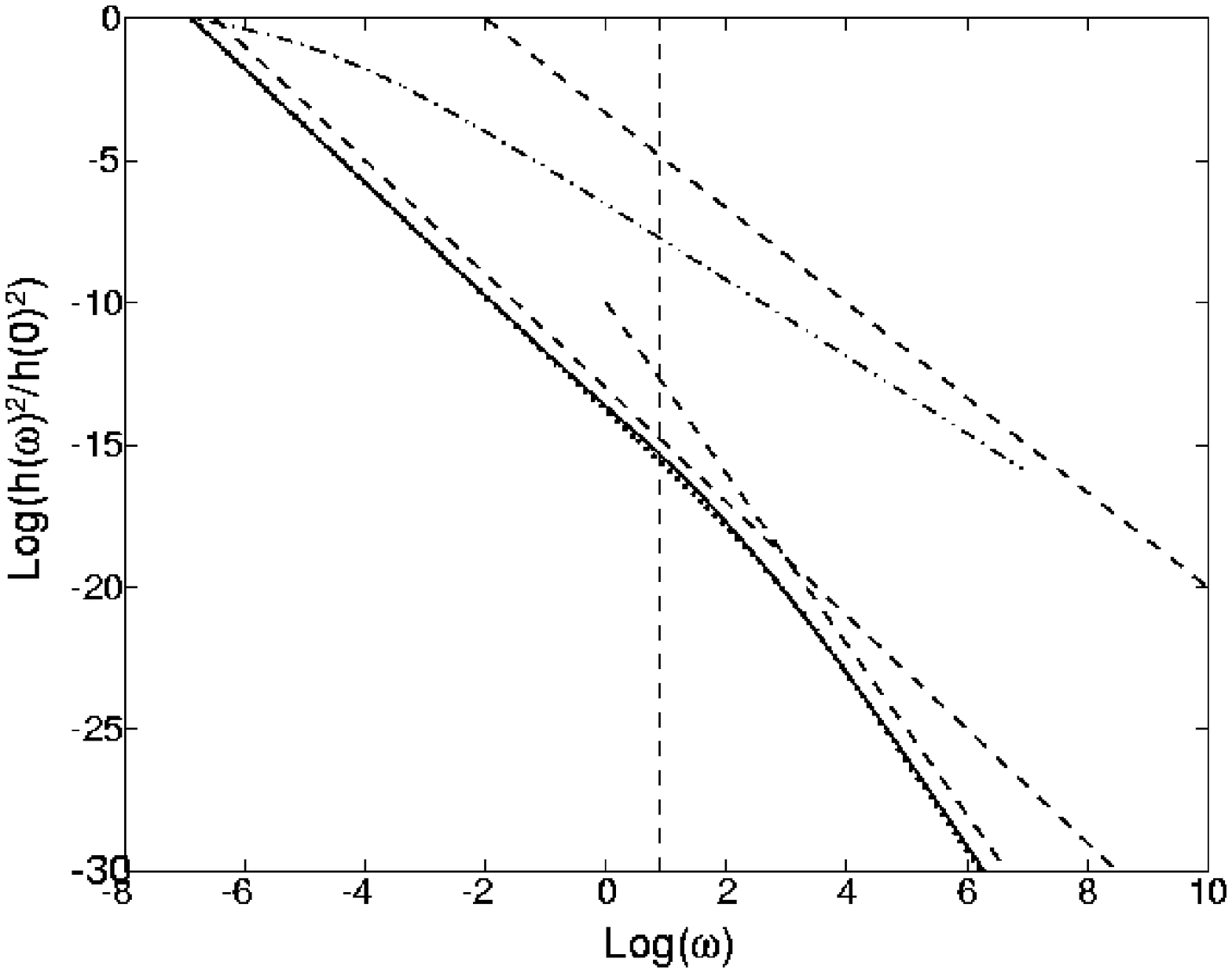}} \caption{}
\end{figure}

\begin{figure}
\centerline{\ \epsfysize 12cm \epsfbox{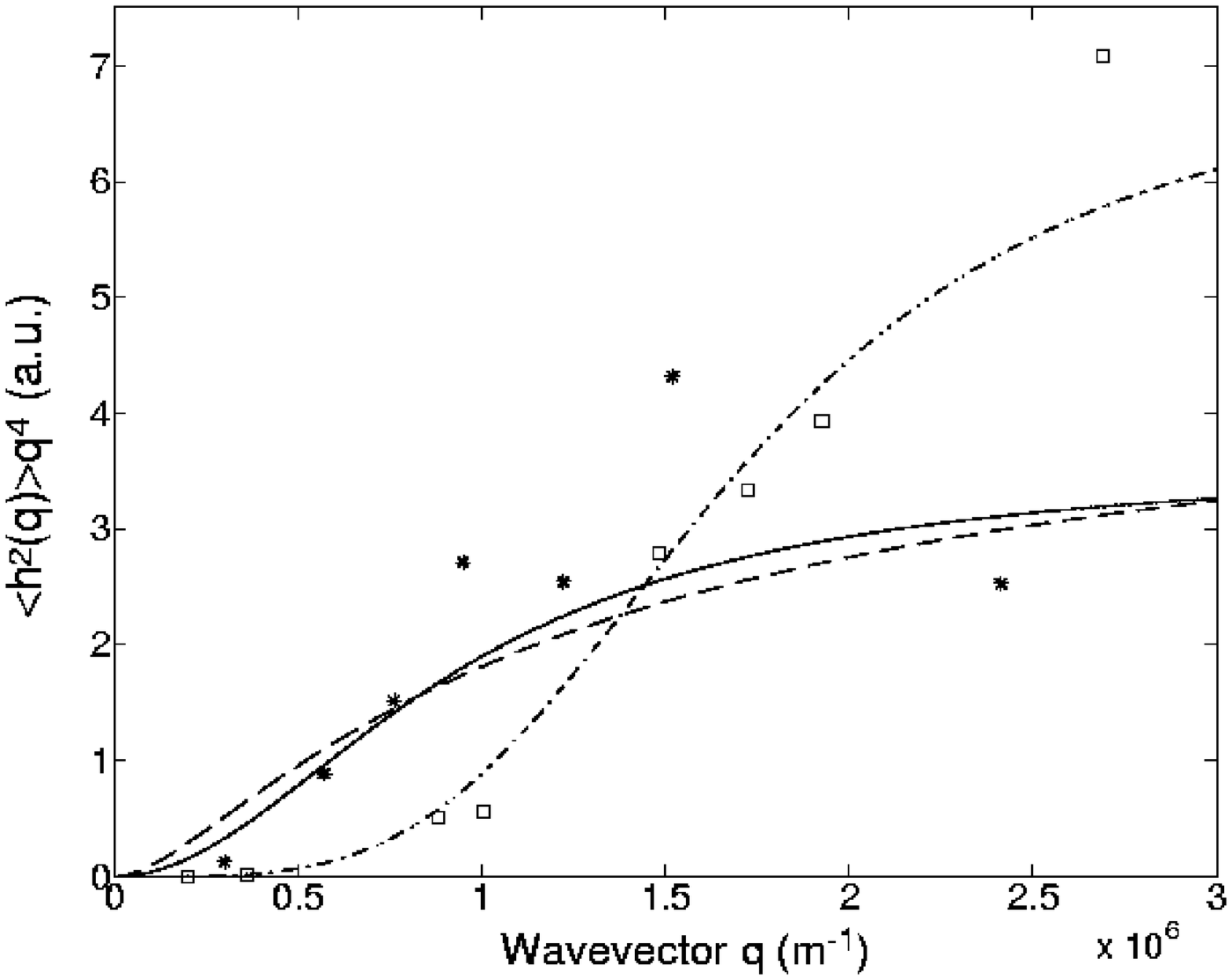}} \centerline{\
\epsfysize 6cm \epsfbox{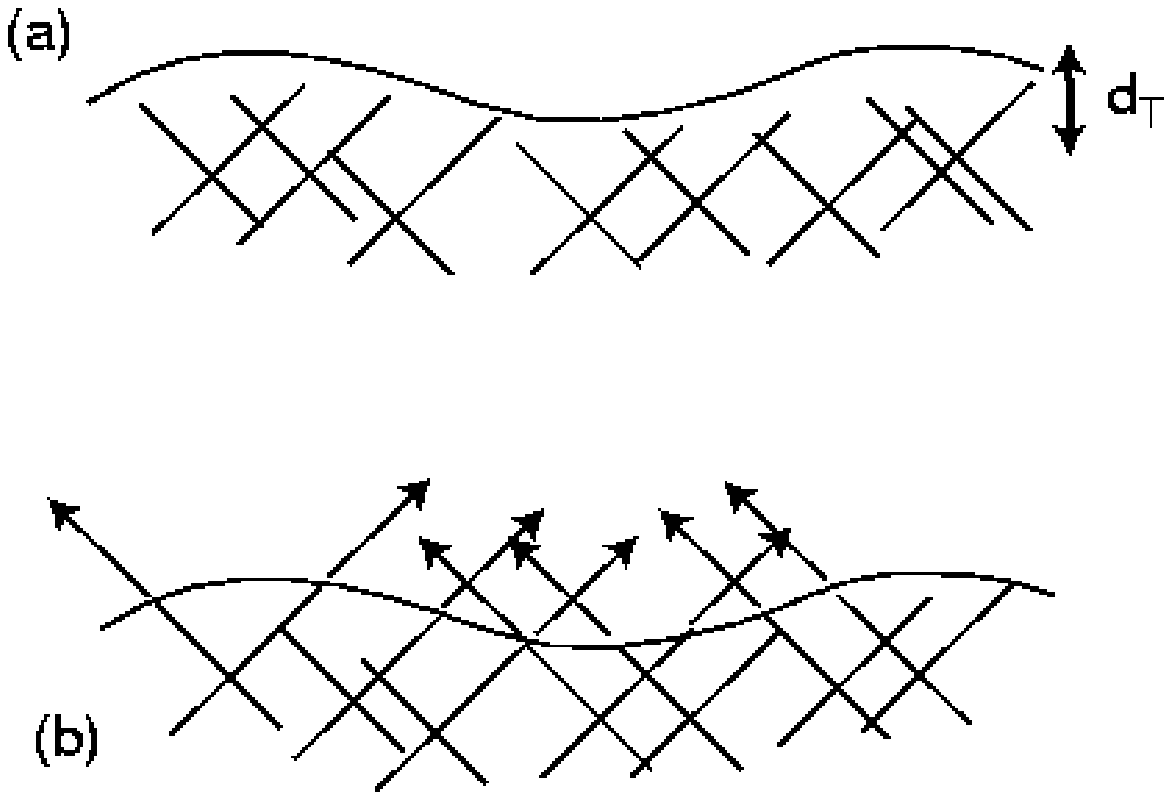}} \caption{}
\end{figure}

\begin{figure}
\centerline{\ \epsfysize 15cm \epsfbox{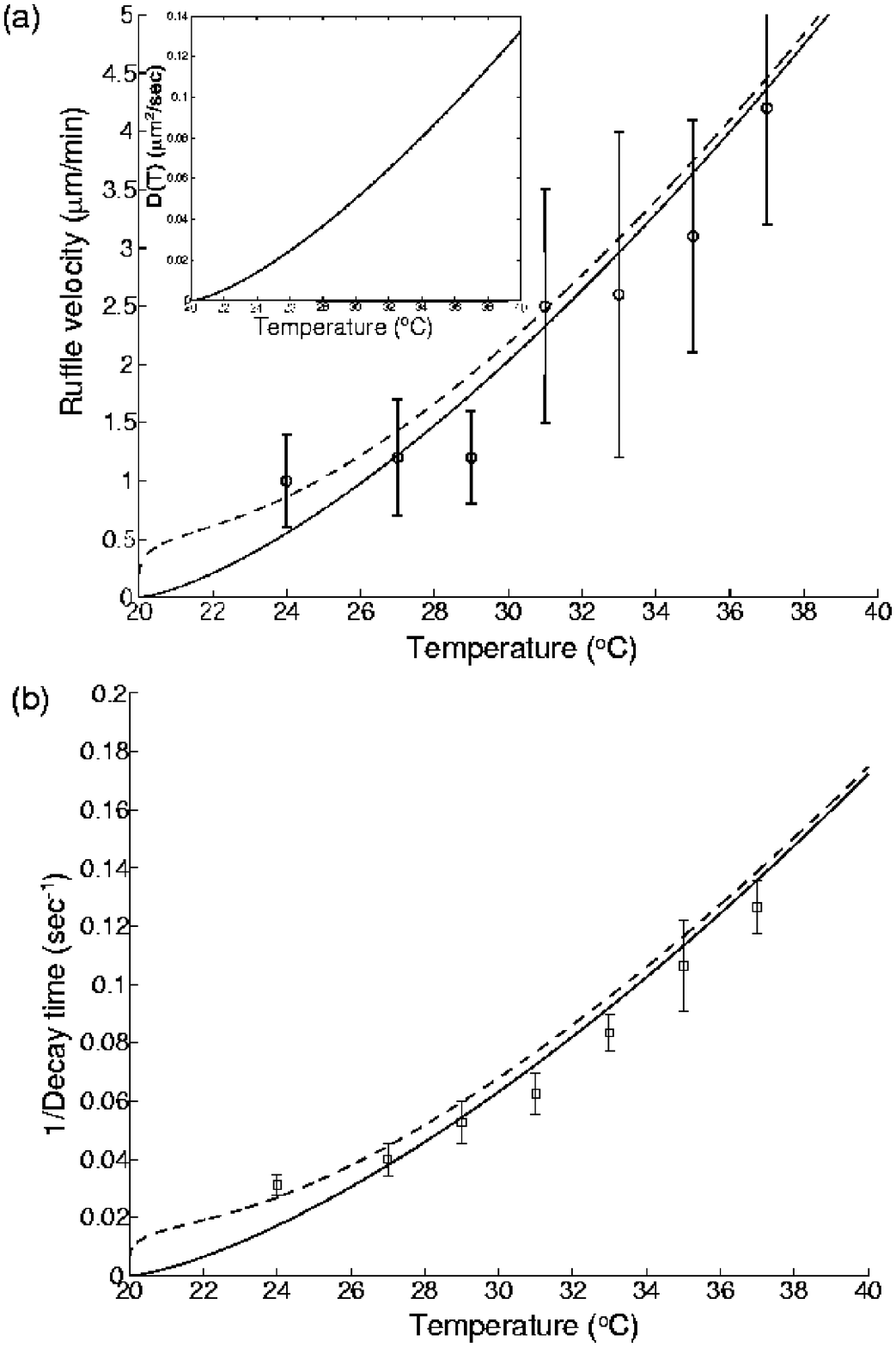}} \caption{}
\end{figure}

\begin{figure}
\centerline{\ \epsfysize 12cm \epsfbox{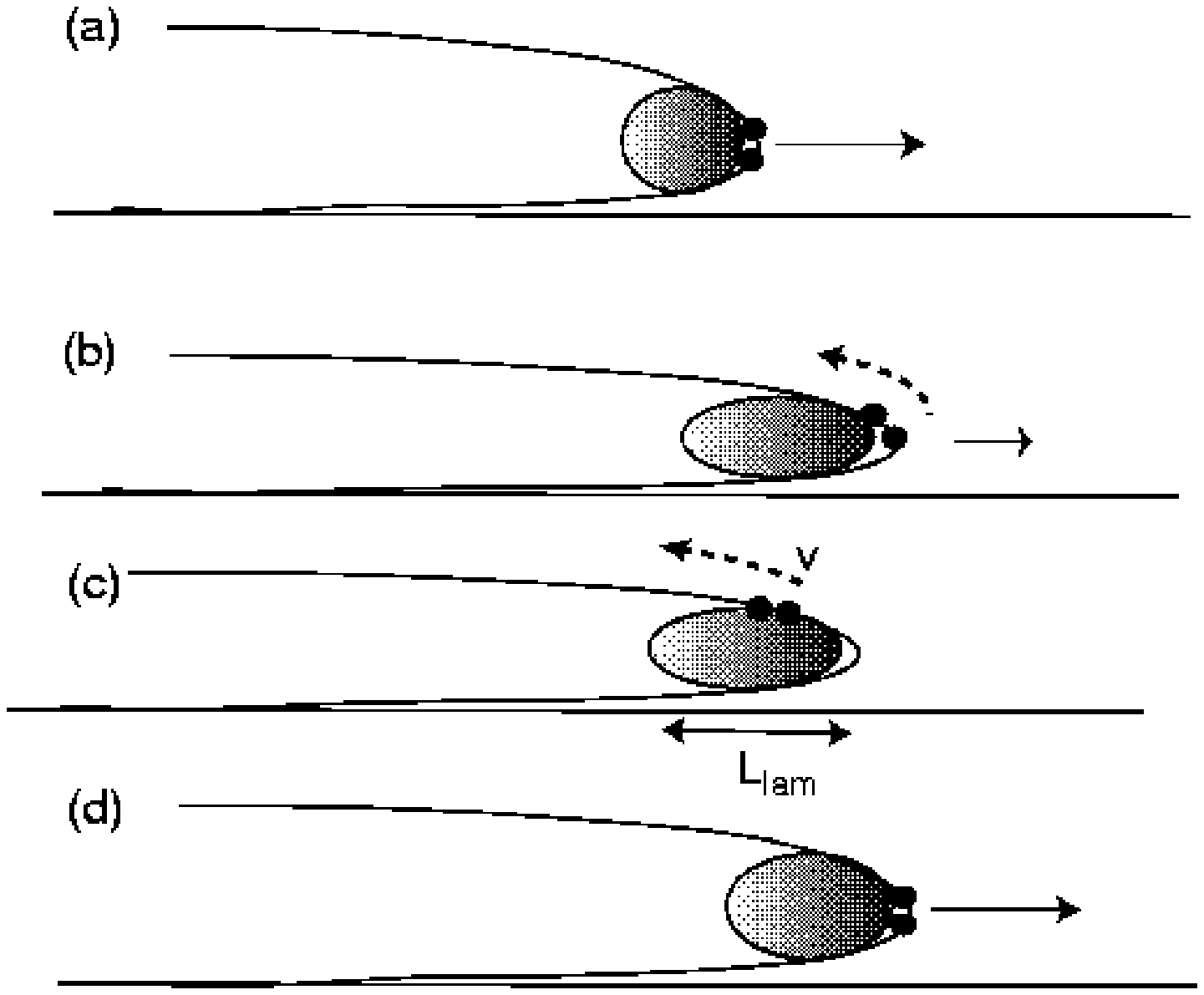}} \caption{}
\end{figure}

\end{document}